\documentclass[a4paper,11pt]{article}
\usepackage{jcappub} 
\usepackage{lineno}
\usepackage{graphicx}
\usepackage{subcaption}
\usepackage[table,xcdraw]{xcolor}
\usepackage{multirow}
\usepackage[normalem]{ulem}
\usepackage{cleveref}

\title{\boldmath Planck constraints on the scale dependence of isotropic cosmic birefringence}

\author[a,b,c,d]{M. Ballardini,}
\author[c,e,a]{A. Gruppuso,}
\author[c,e]{S. Paradiso,}
\author[a,b,f,g]{S.S. Sirletti\footnote{Corresponding author.},}
\author[a,b,c]{and P. Natoli}

\affiliation[a]{Dipartimento di Fisica e Scienze della Terra, Università degli Studi di Ferrara, via Saragat 1, I-44122 Ferrara, Italy}
\affiliation[b]{Istituto Nazionale di Fisica Nucleare, Sezione di Ferrara, via Saragat 1, I-44122 Ferrara, Italy}
\affiliation[c]{Istituto Nazionale di Astrofisica - Osservatorio di Astrofisica e Scienza dello Spazio di Bologna, via Gobetti 101, I-40129 Bologna, Italy}
\affiliation[d]{Department of Physics \& Astronomy, University of the Western Cape, Cape Town 7535, South Africa}

\affiliation[e]{Istituto Nazionale di Fisica Nucleare, Sezione di Bologna, viale Berti Pichat 6/2, I-40127 Bologna, Italy}
\affiliation[f]{Dipartimento di Fisica, Università di Trento, via Sommarive 14, Trento, 38123, , Italy}
\affiliation[g]{Department of Physics, Columbia University, New York, NY 10027, USA}

\emailAdd{mario.ballardini@unife.it}
\emailAdd{alessandro.gruppuso@inaf.it}
\emailAdd{simone.paradiso@inaf.it}
\emailAdd{salvatoresamuele.sirletti@unife.it}
\emailAdd{paolo.natoli@unife.it}

\abstract{The rotation of the linear polarisation plane of photons during propagation, also known as cosmic birefringence, is a powerful 
probe of
parity-violating extensions of standard electromagnetism. 
Using \textit{Planck} legacy data, we confirm
previous estimates of the isotropic birefringence angle, finding $\beta \simeq 0.30 \pm 0.05$ [deg] at 68\% CL, not including the systematic error from the instrumental polarisation angle. 
If this is a genuine signal, it could be explained by theories of Chern--Simons-type coupled to electromagnetism, which could lead to a harmonic scale-dependent birefringence signal, if the hypothesis of an ultra-light (pseudo) scalar field does not hold. To investigate
these models, we pursue 
two complementary approaches: first, we fit the birefringence angle estimated at different multipoles, $\beta_{\ell}$,
with a power-law model and second, we perform a non-parametric Bayesian reconstruction of it.
Both methods yield results consistent with a non-vanishing constant birefringence angle. The first method shows no significant dependence on the harmonic scale (up to $1.8\sigma$ CL), while the second method demonstrates that a constant model is favored by Bayesian evidence.
This conclusion is robust across all four published \textit{Planck} CMB solutions.
Finally, we forecast that upcoming CMB observations by Simons Observatory, LiteBIRD and a wishful CMB-Stage 4 experiment could reduce current uncertainties by a factor of approximately 7.
}

\begin{document}
\maketitle
\flushbottom

\section{Introduction}
\label{sec:intro}

The cosmic birefringence (CB) effect, 
also known as cosmic polarisation rotation (CPR), is the rotation of the linear polarisation plane of photons in vacuum \cite{Carroll:1989vb,Carroll:1991zs,Harari:1992ea,Carroll:1998zi}. Parity-violating extensions of standard electromagnetism can produce this effect; 
the most notable example
is Chern-Simons (ChS) theory \cite{Carroll:1989vb}. 
Note that this rotation is expected to be 
zero in the standard Maxwellian scenario, see also Refs.~\cite{Ni:1977zz,Ni:2016dvq}. 
Since the cosmic microwave background (CMB) radiation is (1) linearly polarised to a few percent level due to Thomson scattering at the last scattering surface, 
and (2) the most distant known source of linear polarisation in nature, it has become customary to constrain the CB through observations of the CMB polarisation \cite{Komatsu:2022nvu}. 

From an observational point of view, the CB eﬀect induces a mixing between $E$- and $B$-mode polarisations, which in turn switches on correlations that would not otherwise exist. Specifically, CB affects the observed CMB angular power spectra (APS), $C_\ell^{X}$ with $X=T,E,B$, as follows \cite{Lue:1998mq,Feng:2004mq,Liu:2006uh}
\begin{align}
    C_\ell^{TT,\text{obs}} &= C_\ell^{TT} \,, \label{TTobs}\\
    C_\ell^{TE,\text{obs}} &= C_\ell^{TE}\cos(2\beta) \,, \label{TEobs}\\
    C_\ell^{TB,\text{obs}} &= C_\ell^{TE}\sin(2\beta) \,, \label{TBobs}\\
    C_\ell^{EE,\text{obs}} &= C_\ell^{EE}\cos^2(2\beta) + C_\ell^{BB}\sin^2(2\beta) \,, \label{EEobs}\\
    C_\ell^{BB,\text{obs}} &= C_\ell^{BB}\cos^2(2\beta) + C_\ell^{EE}\sin^2(2\beta) \,, \label{BBobs}\\
    C_\ell^{EB,\text{obs}} &= \frac{1}{2} \left(C_\ell^{EE} - C_\ell^{BB}\right)\sin(4\beta) \,, \label{EBobs}
\end{align}
where $C_\ell^{X,\text{obs}}$ are the observed CMB APS (i.e. those 
containing the effect), while on the right-hand side there are the unrotated ones, $C_\ell^{X}$, that would be observed 
if the birefringence angle $\beta$ 
was null. It is important to note that \cref{TEobs,TBobs,EEobs,BBobs,EBobs} are strictly valid under the assumption that no parity-violation signals are present at the last scattering surface and that the birefringence angle $\beta$ is constant, see Refs.~\cite{Finelli:2008jv,Fedderke:2019ajk,Galaverni:2023zhv} for generalisations. As expected, note also that $C_\ell^{TT,\text{obs}}$ is not sensitive to the CB effect, 
as this only affects the CMB polarisation.

The \textit{Planck} collaboration constrained the CB angle to be $\beta = 0.31 \pm 0.05\,\mbox{(stat)} \pm 0.28 \,\mbox{(sys)}$ [deg], where the systematic uncertainty, due to the instrumental polarisation angle (henceforth denoted as $\alpha$) uncertainty, was found to dominate the total error budget \cite{Planck:2016soo}. Specifically, $\alpha$ affects the CMB APS in the same way as described in~\cref{TEobs,TBobs,EEobs,BBobs,EBobs} \cite{Keating:2012ge}; therefore, a degeneracy exists between this systematic effect and the cosmological CB effect (see the discussion below).
To break this degeneracy, a new technique, now known as the Minami-Komatsu method, was proposed in Refs.~\cite{Minami:2019ruj,Minami:2020fin} and applied to \textit{Planck} data in Ref.~\cite{Minami:2020odp} and Ref.~\cite{Diego-Palazuelos:2022dsq,Diego_Palazuelos_2023}, suggesting a detection of $\beta$ around $0.3$~[deg] with a statistical significance close to $3 \sigma$ confidence level (CL). 
In particular, WMAP and \textit{Planck} data were analysed jointly in Ref.~\cite{Eskilt:2022cff}, leading to the exclusion of $\beta = 0$ at a significance level of 3.6$\sigma$. See also Ref.~\cite{Gruppuso:2025ywx} for a review of recent constraints on the CB effect.
Furthermore, we note that the ACT collaboration has recently suggested an estimate of the birefringence angle, $\beta_{\mathrm{ACT}} = 0.20 \pm 0.08$~[deg], based on DR6 data~\cite{ACT:2025fju}. In their analysis, 
they make use of a detailed optical modelling.\footnote{While this ACT constraint is noteworthy, it is obtained without a dedicated component separation step, and also does not account for pointing uncertainty, which is of the order of $0.03$~[deg],  
see again Ref.~\cite{ACT:2025fju}.}
See also Ref.~\cite{Sullivan:2025btc} for recent estimates of the CB angle from \textit{Planck} Public Data Release 4 data employing a pixel-based estimator.

These estimates are typically interpreted by extending standard electromagnetism.\footnote{See also Ref.~\cite{COMPACT:2024cud} for an interpretation based on non-trivial spatial topology, which leaves standard electromagnetism unchanged.}
In the context of ChS extensions \cite{Carroll:1991zs}, the photon is coupled to a (pseudo) scalar field $\chi$, also called axion or axion-like particle (ALP) depending on the specific potential of the field $\chi$, through the following interaction term
\begin{equation}
    {\cal L}_\text{int} = g \, \chi \, \tilde F_{\mu \nu} \, F^{\mu \nu} \,,
\end{equation}
where $g$ is a coupling constant with dimensions of $[\text{energy}]^{-1}$, $F_{\mu \nu}$ is the electromagnetic tensor and $\tilde F_{\mu \nu}$ is its dual.
It can be show that the \cref{TEobs,TBobs,EEobs,BBobs,EBobs} are recovered by the ChS theories when $\chi$ is almost constant in time during the recombination epoch \cite{Nakatsuka:2022epj}. This occurs when the mass of the field is sufficiently small for a fixed coupling constant $g$. This qualitative behaviour appears to be largely independent of the potential considered for $\chi$ \cite{Greco:2024oie}. 
On the contrary, if the mass is larger, $\chi$ exhibits a time-evolution (characterized by a decay followed by damped oscillations) that modifies these relations.
As a consequence, in ChS extensions of standard electromagnetism, the degeneracy between $\alpha$ and $\beta$ is exact when the mass of the ALP is ultra-light, otherwise such a degeneracy is broken, see also Refs.~\cite{Fujita:2020ecn,Yin:2023srb,Kochappan:2024jyf}. In Ref.~\cite{Greco:2022xwj} the dependence on the mass of the ALP is studied through a related effect, the anisotropic birefringence.

From an observational perspective, the degeneracy is maximised when $\beta$ is independent of the harmonic scale. Previous analyses of \textit{Planck} measurements \cite{Eskilt:2023nxm} suggest that this scenario is currently favoured by the data.
Similar conclusions are obtained in \cite{Namikawa:2025sft}, where constraints on the ALPs parameters are derived from the full shape of the \textit{Planck} \textit{EB} spectrum.
In this work, we propose a new test to assess whether $\beta$ is indeed independent of the harmonic scale. 
We use \textit{Planck} Public Data Release 3 (henceforth PR3) \cite{Planck:2018nkj} to estimate the birefringence angles at different harmonic scales, hereafter denoted as $\beta_{\ell}$, and we test their consistency across the entire harmonic range.\footnote{Note that the goal of this paper is not to address the degeneracy between $\beta$ and $\alpha$ directly. However, the proposed test may help 
to distinguish between the two.}
This is performed in two different ways, as detailed below:
\begin{enumerate}
\item In the first approach, we fit the estimated angles at different angular scales using a two-parameter power-law model, $(\beta_0, n)$, defined as $\beta_{\ell} = \beta_0 \left( \ell / \ell_0 \right)^n$, where $\ell_0$ is a suitable harmonic pivot scale.\footnote{As will become clear in the following sections, the power-law functional is actually employed in a slightly different form, where the harmonic dependence is 
expressed in terms of band powers.}
In this approach, the power $n$ is the key parameter, as it quantifies the consistency of a constant birefringence angle across the entire harmonic range under consideration.
\item In the second approach, we fit $\beta_{\ell}$ using a non-parametric Bayesian reconstruction for a given number of moving knots
\cite{Handley:2015fda,Handley:2015vkr,Raffaelli:2025kew}. In this case, the best-fit curve is statistically evaluated against constant behavior in $\ell$.
\end{enumerate}

We anticipate that the first method indicates that the CB angle is independent of the harmonic scale up to $1.8\sigma$ CL. A consistent result is obtained with the second approach, i.e. the non-parametric method, where a constant birefringence angle is preferred in terms of Bayesian evidence against more flexible models that require a larger number of parameters. 

The paper is organised as follows. In \cref{sec:descriptiondata}, we describe the \textit{Planck} data and the corresponding simulations used in our analyses. In particular, we compute the CMB spectra for both simulations and data, and assess the statistical significance of the deviation of the $EB$ spectrum from zero.
In \cref{sec:test}, we test the harmonic-scale dependence of the CB angle. Specifically, in \cref{sec:estimatesbetaell}, we estimate the birefringence angles in different harmonic ranges, $\beta_{\ell}$, and characterise their covariance using simulations. These $\beta_{\ell}$ values are then fitted with a power-law model in \cref{sec:power-law}, and interpolated using a non-parametric approach in \cref{sec:Nonparametric}. In \cref{sec:forecasts} we provide forecasts for future CMB experiments. Finally, in \cref{sec:conclusions}, we present our conclusions, along with a statistical assessment of the results.

\section{Planck data and simulations}
\label{sec:descriptiondata}

In this work, we employ \textit{Planck} PR3 data and simulations~\cite{Planck:2018nkj}. Specifically, we make use of the official \textit{Planck} CMB maps obtained through four component-separation (CS) methods: \texttt{Commander}~\cite{Eriksen:2005dr, Eriksen:2007mx}, \texttt{NILC}~\cite{Delabrouille:2008qd}, \texttt{SEVEM}~\cite{Martinez-Gonzalez:2003abe}, and \texttt{SMICA}~\cite{Delabrouille:2002kz, Cardoso:2008qt}. These methods process multi-frequency observations to extract the CMB signal, cleaned from foreground emissions~\cite{Leach:2008fi, Planck:2018yye}.
In order to statistically characterize the data, we exploit the official \textit{Planck} PR3 simulations.
For each CS method, the simulation set consists of 1000 CMB Monte Carlo simulations generated using the \textit{Planck} $\Lambda$CDM best-fit model (also referred to as the fiducial model) \cite{Planck:2018yye} including a beam modeling,\footnote{In polarization this is equivalent to an effective Gaussian beam of $5$~arcmin, see Refs.~\cite{Bortolami:2022whx,2010A&A...520A...1T}.} along with 300 noise simulations for the first half-mission (HM1) and 300 for the second half-mission (HM2). The \textit{Planck} PR3 simulations include residuals from several systematic effects, such as beam leakage, analog-to-digital converter (ADC) non-linearities, thermal fluctuations, and others \cite{Planck:2015txa,Planck:2018lkk}. 
All product used here are publicly available on the ESA \textit{Planck} Legacy Archive\footnote{\url{http://pla.esac.esa.int/pla}} and are used at $N_{\text{side}}=2048$, where $N_{\text{side}}$ is a parameter of the Hierarchical Equal Area isoLatitude Pixelization of a sphere (\texttt{HEALPix}) pixelisation scheme \cite{Gorski:2004by},\footnote{\url{https://sourceforge.net/projects/healpix/}} which is related to the number of pixels, $N_{\text{pix}}$, through $N_{\text{pix}}= 12 N_{\text{side}}^2$.

\subsection{Angular power spectra}
\label{angularpowerspectra}

\begin{figure}[!ht]
    \centering
    \includegraphics[width=\textwidth]{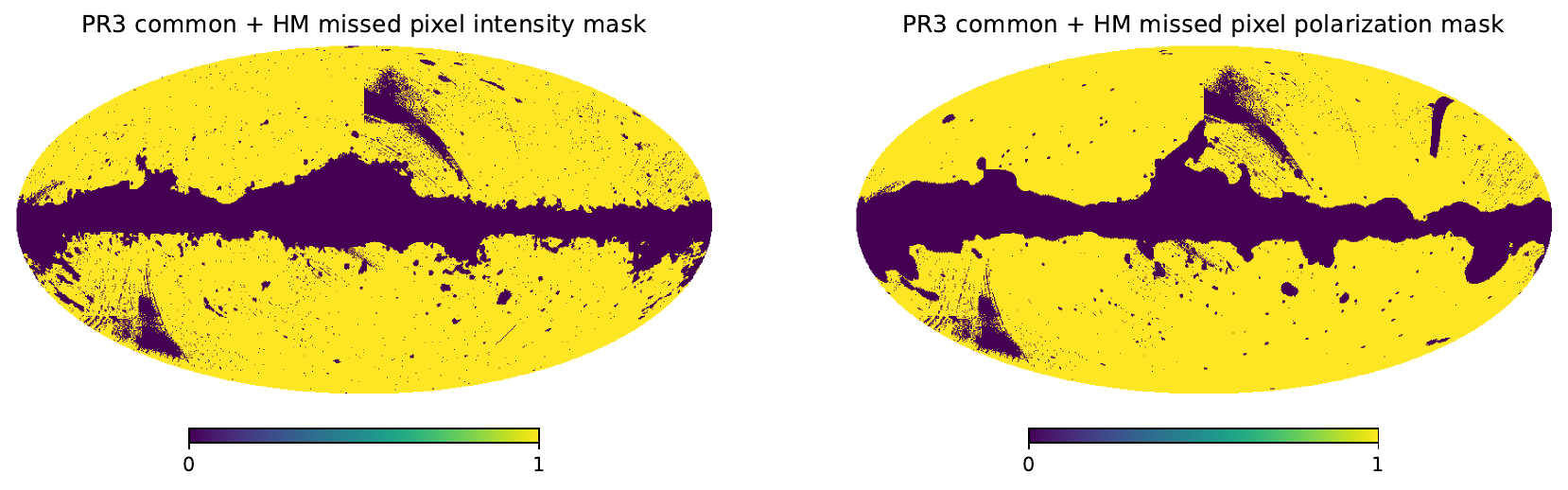}
    \caption{Masks used in this work. Both the intensity (left panel) and polarization (right panel) masks  
    have an effective sky fraction of $f_{\text{sky}} \simeq 75 \%$.}
    \label{fig:combined_mask_maps}
\end{figure}

\begin{figure}[!ht]
    \centering
    \begin{minipage}{0.49\textwidth}
        \centering
        \includegraphics[width=\textwidth]{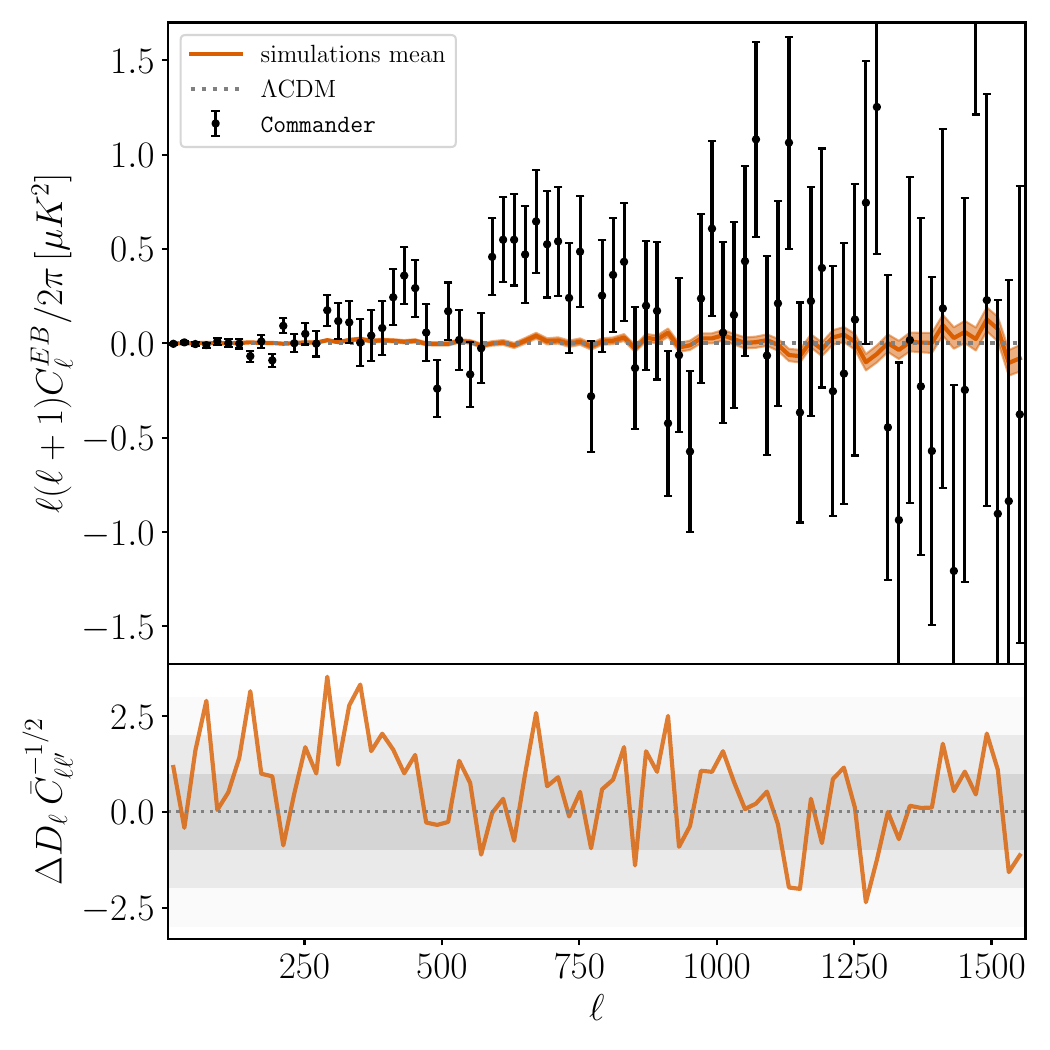}
    \end{minipage}
    \hfill
    \begin{minipage}{0.49\textwidth}
        \centering
        \includegraphics[width=\textwidth]{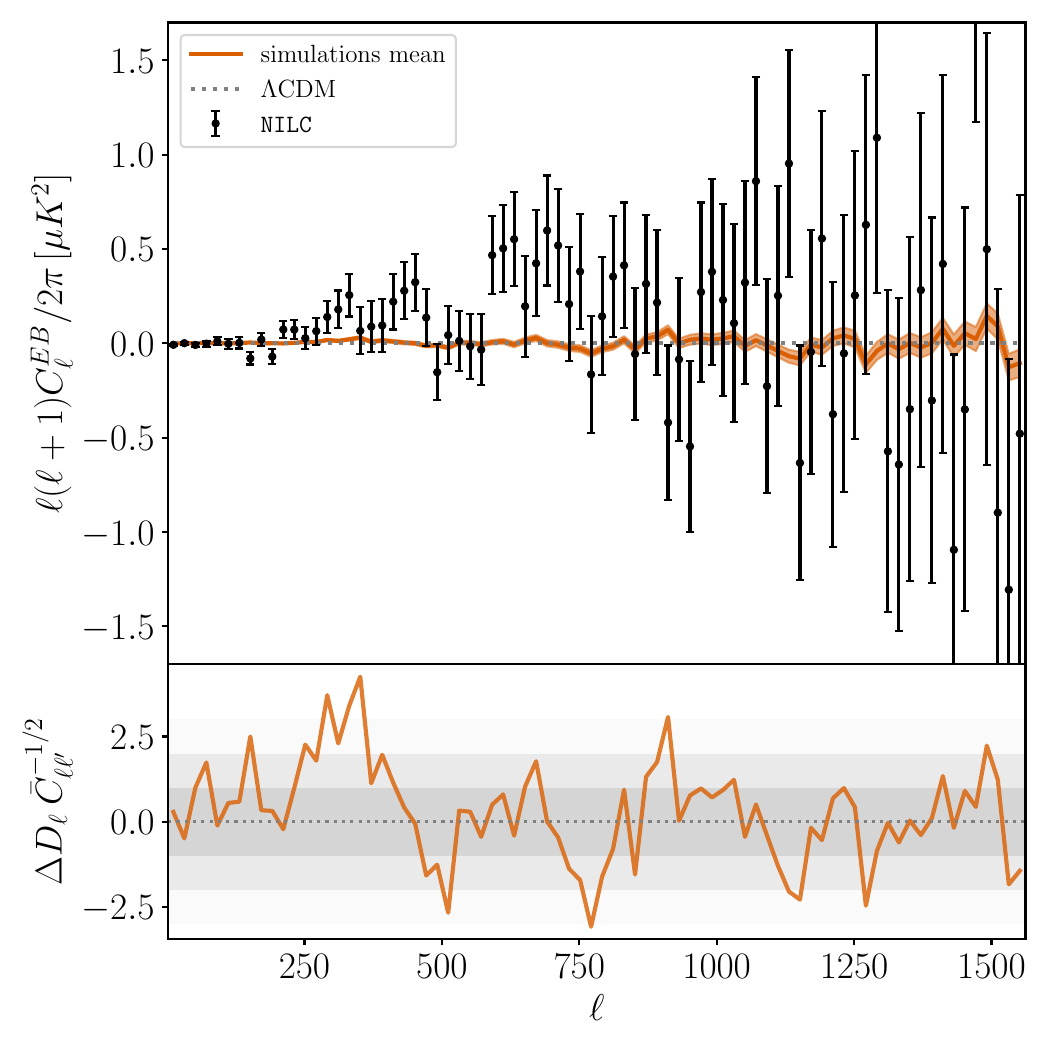}
    \end{minipage}
    \hfill
    \begin{minipage}{0.49\textwidth}
        \centering
        \includegraphics[width=\textwidth]{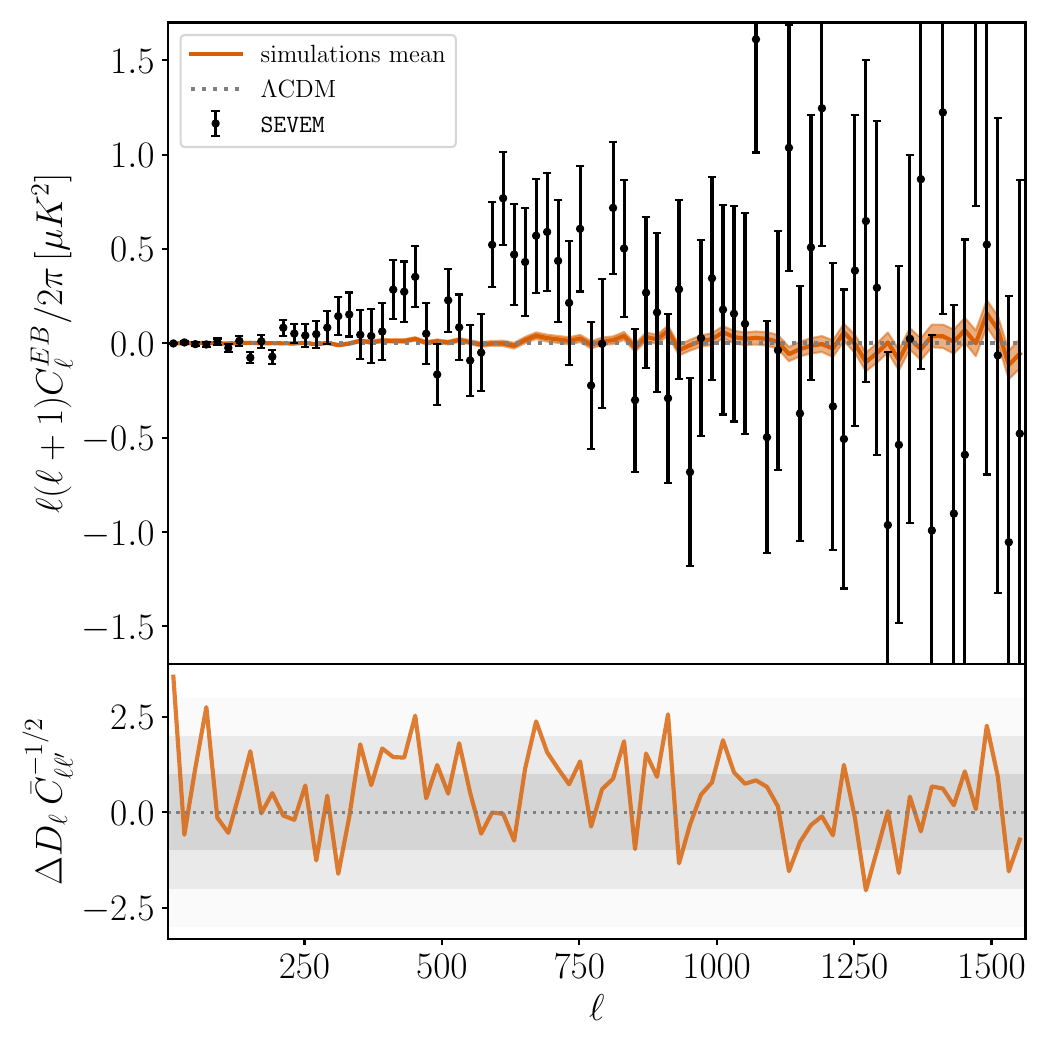}
    \end{minipage}
    \hfill
    \begin{minipage}{0.49\textwidth}
        \centering
        \includegraphics[width=\textwidth]{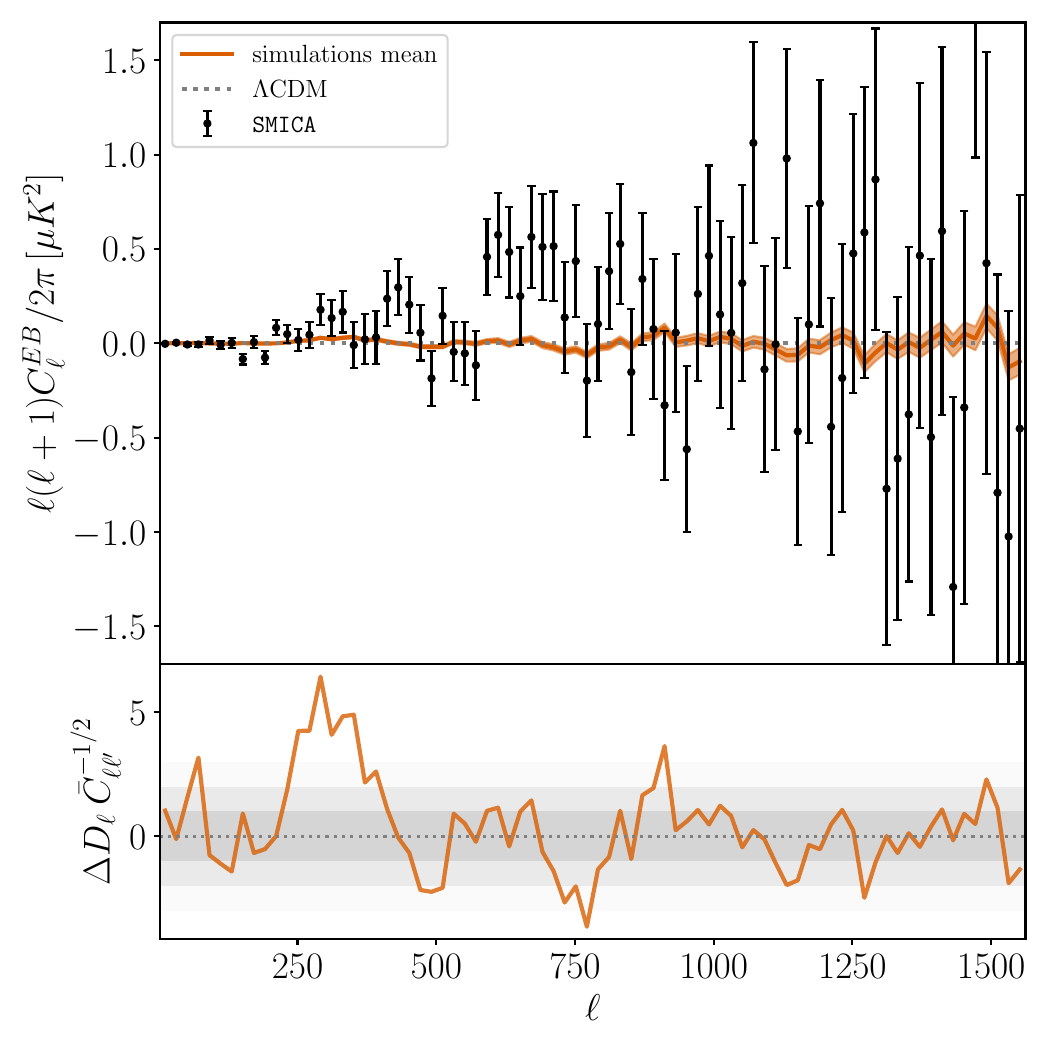}
    \end{minipage}
    \caption{$\mathcal{D}^{EB}_\ell \equiv \ell (\ell+1) C^{EB}_\ell / 2\pi$ [$\mu K^2$] versus $\ell$ from \textit{Planck} PR3 data binned with $\Delta \ell = 20$. The panels are organised by CS method: upper left panel for \texttt{Commander}, upper right for \texttt{NILC}, lower left for \texttt{SEVEM}, and lower right for \texttt{SMICA}. In each panel, the upper part shows the APS with error bars (black), the $\Lambda$CDM fiducial spectrum (dotted gray), and the mean with one $\sigma$ dispersion of the mean from the simulations (orange). The lower part shows the fluctuation of the mean of the simulations relative to the fiducial spectrum in units of the standard deviation of the mean.}
    \label{fig:EB_validation_and_spectra}
\end{figure}

\begin{figure}[!ht]
    \centering
    \begin{minipage}{0.49\textwidth}
        \centering
        \includegraphics[width=\textwidth]{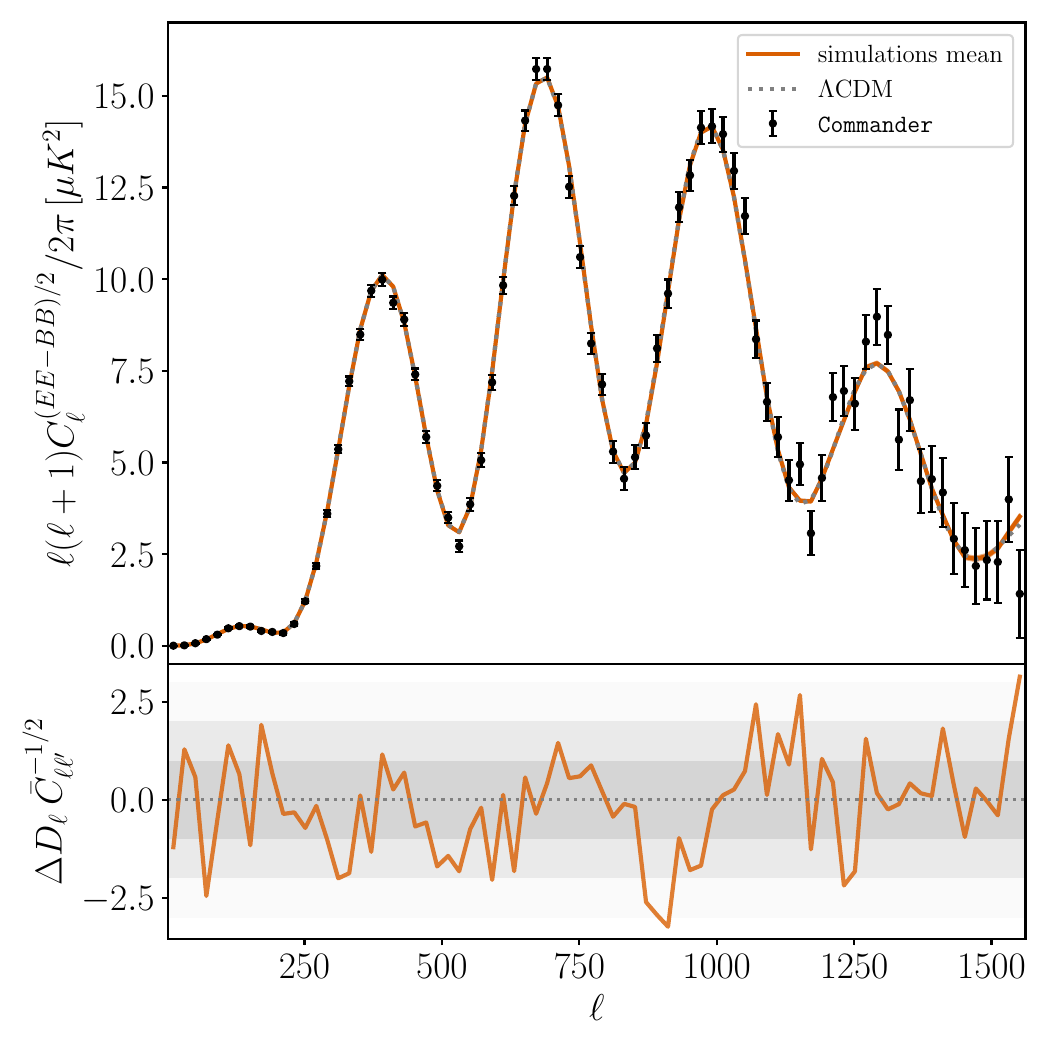}
    \end{minipage}
    \hfill
    \begin{minipage}{0.49\textwidth}
        \centering
        \includegraphics[width=\textwidth]{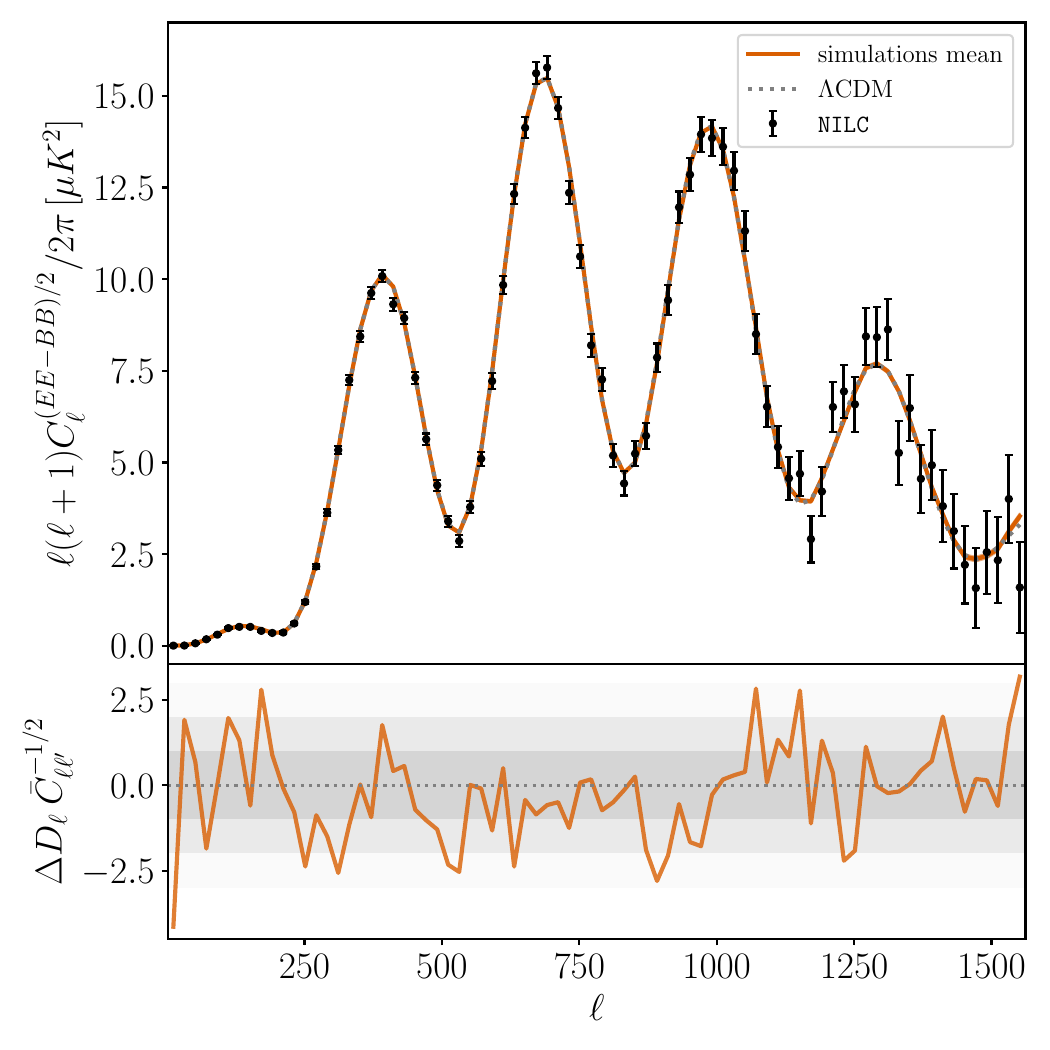}
    \end{minipage}
    \hfill
    \begin{minipage}{0.49\textwidth}
        \centering
        \includegraphics[width=\textwidth]{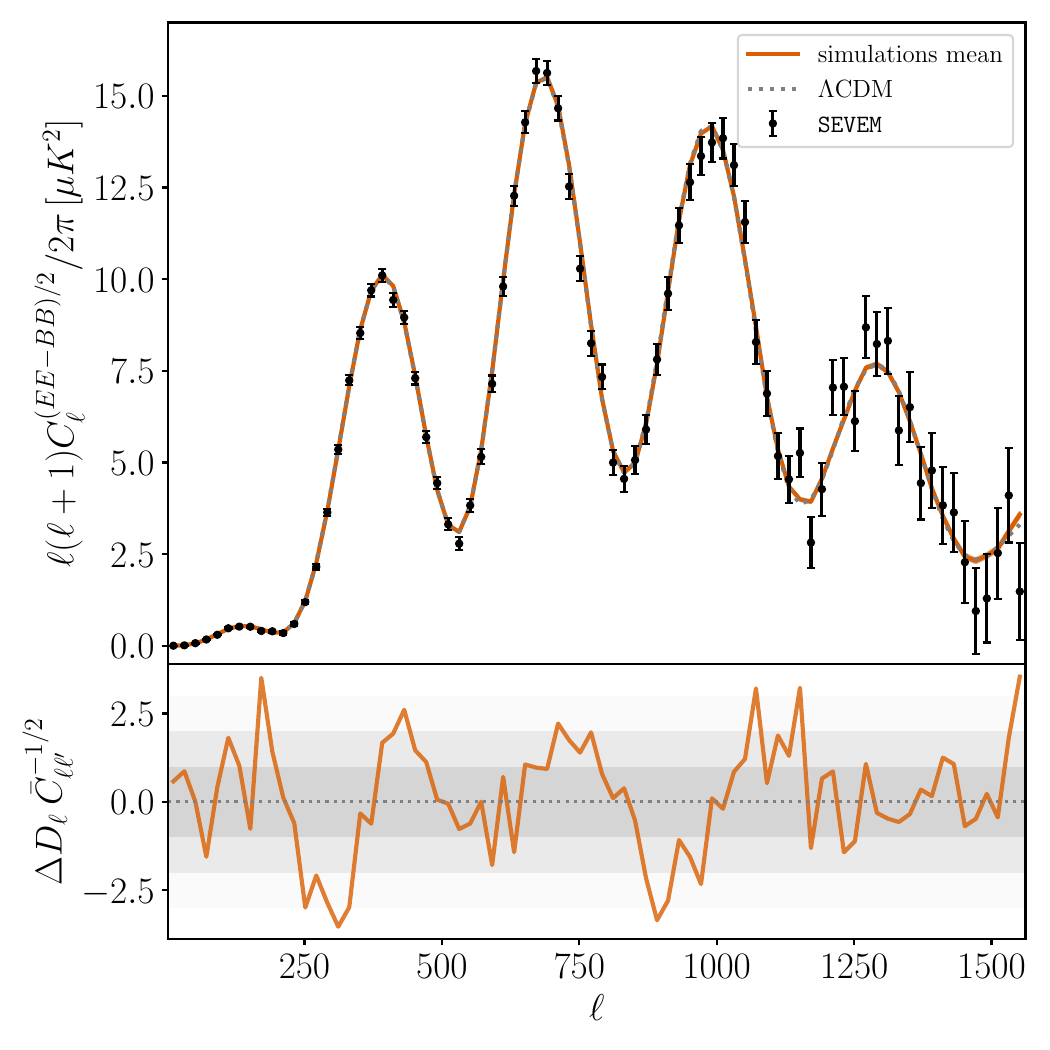}
    \end{minipage}
    \hfill
    \begin{minipage}{0.49\textwidth}
        \centering
        \includegraphics[width=\textwidth]{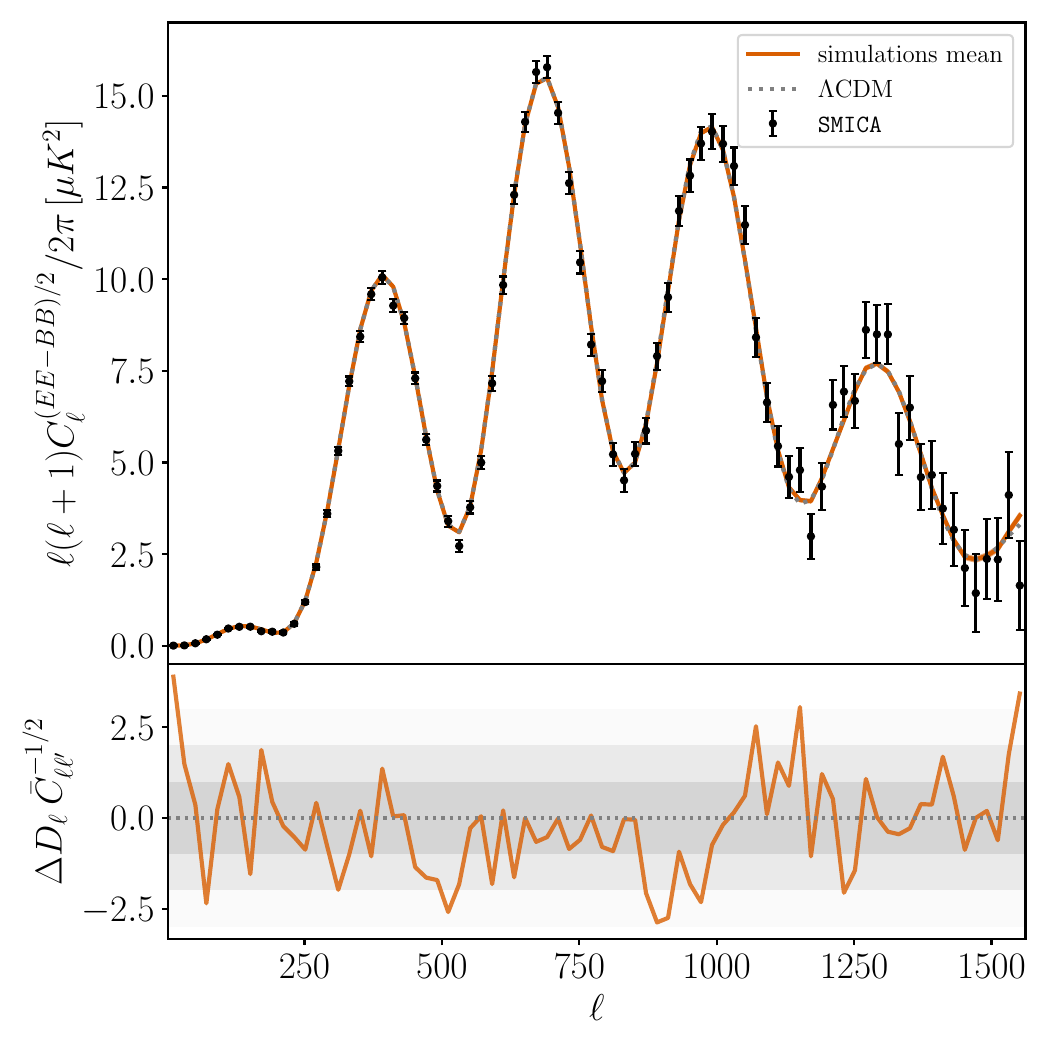}
    \end{minipage}
    \caption{$\mathcal{D}^{(EE-BB)/2}_\ell \equiv \ell (\ell+1) [(C^{EE}_\ell - C^{BB}_\ell)/2] / 2\pi$ [$\mu K^2$] versus $\ell$ from \textit{Planck} PR3 data binned with $\Delta \ell = 20$. The panels are organized by CS method: upper left panel for \texttt{Commander}, upper right for \texttt{NILC}, lower left for \texttt{SEVEM}, and lower right for \texttt{SMICA}. In each panel, the upper part shows the APS with error bars (black), the $\Lambda$CDM fiducial spectrum (dotted grey), and the mean with one $\sigma$ dispersion of the mean from the simulations (orange). The lower part shows the fluctuation of the mean of the simulations relative to the fiducial spectrum in units of the standard deviation of the mean.}
    \label{fig:EEmBBo2_validation_and_spectra}
\end{figure}

For each CS method, we build two data splits by co-adding the 300 HM1 and the 300 HM2 noise simulations with the first 300 CMB simulations. These splits are then used to estimate the CMB APS in cross-mode, which is known to provide auto-spectra that are less sensitive to noise details and residual systematic effects
\cite{Planck:2018yye}.
The APS are computed using the \texttt{NaMaster} Python package \cite{Alonso:2018jzx},\footnote{\url{https://namaster.readthedocs.io}} which implements the so-called pseudo-$C_{\ell}$ approach \cite{Hivon:2001jp,Polenta:2004qs}. This computation 
takes into account the \textit{Planck} PR3 common masks for intensity and polarisation, as well as the masks covering unobserved pixels in HM1 and HM2, see \cref{fig:combined_mask_maps}. No mask apodization or purification \cite{Lewis:2001hp, Bunn:2002df, Grain:2009wq} is applied. 
See also Ref.~\cite{Bortolami:2022whx} though employing different sky coverage and band powers.
\Cref{fig:EB_validation_and_spectra,fig:EEmBBo2_validation_and_spectra} display $\mathcal{D}^{EB}_\ell \equiv \ell (\ell+1)C^{EB}_{\ell}/ 2\pi$ and $\mathcal{D}^{(EE-BB)/2}_\ell \equiv \ell (\ell+1) \left[ (C^{EE}_{\ell} - C^{BB}_{\ell})/2 \right] /2\pi$, respectively. This specific APS choice is motivated by the analysis pipeline described in \cref{sec:estimatesbetaell}.
Each of these figures contains four panels, one for each CS method, presenting the binned APS with bandpowers of $\Delta \ell = 20$ up to a maximum multipole $\ell_{\rm max} \simeq 1550 $. In each panel, we illustrate the data APS with error bars (estimated from the simulations) in black, the $\Lambda$CDM fiducial spectrum in dotted gray, and the mean with one $\sigma$ dispersion from the simulations in orange. 
The lower part of each panel shows how well the simulations are compatible with the expected fiducial APS, denoted by $\mathcal{D}^{X,{\rm fid}}_\ell$, where $X=\{EB,(EE-BB)/2\}$. To achieve this, we plot \textit{standardised} APS, defined as:
$$
\Delta \mathcal{D}^X_\ell \cdot (\bar{\mathrm{C}}^X)^{-1/2}_{\ell\ell'} \,,
$$
where, $\Delta \mathcal{D}^X_\ell = \langle \mathcal{D}^X_\ell - \mathcal{D}^{X,{\rm fid}}_\ell \rangle$ represents the mean difference from the fiducial spectrum. 
The term $(\bar{\mathrm{C}}^{\rm X})^{-1/2}_{\ell\ell'}$ is the Cholesky decomposition of the inverse of the mean covariance, $\bar{\mathrm{C}}^X_{\ell\ell'}$. The latter is calculated as $\mathrm{C}^X_{\ell\ell'}/300$, where $\mathrm{C}^X_{\ell\ell'}$ is the covariance of $X$.
These quantities are assumed to be approximately normal distributed with zero mean and unitary standard deviation at each harmonic bandpower, and therefore expected to lie mostly within $\pm 3\sigma$.
The covariances $\mathrm{C}^X_{\ell\ell'}$ are computed using an analytical approach provided by the \texttt{NaMaster} code, which includes the fluctuations of signal and noise, as well as the mode coupling induced by the sky masking, see again Ref.~\cite{Alonso:2018jzx}.

The lower panels of \cref{fig:EB_validation_and_spectra,fig:EEmBBo2_validation_and_spectra} show that the simulation average deviates by less than 3.5$\sigma$ from the fiducial for all the considered CS methods. This indicates that the residuals present in the official \textit{Planck} simulations do not exhibit any significant systematic behaviour. The sole outstanding exception is observed for \texttt{SMICA} in $\mathcal{D}^{EB}_\ell$ around $\ell \simeq 300$, where a fluctuation of $\sim 5\sigma$ 
is found. However, such a deviation corresponds to only around $30\%$ of the APS statistical uncertainty 
at those angular scales. Therefore, we do not expect this excess to significantly impact our results. Moreover, to reinforce this point, the corresponding \texttt{SMICA} estimates are well consistent with those derived from the other CS methods, even around $\ell \simeq 300$ where the simulations of the other CS methods do not exhibit any significant deviation from the expected statistical fluctuations.

In summary, we find excellent consistency of the APS estimates for both $\mathcal{D}^{EB}_\ell$ and $\mathcal{D}^{(EE-BB)/2}_\ell$ at better than $2\sigma$ level across all the CS methods. Moreover, a harmonic $\chi^2$-analysis\footnote{$\chi^2=\sum_{\ell,\ell'}\mathcal{D}^X_\ell {\mathrm{C}^{\rm X}_{\ell\ell'}}^{-1} \mathcal{D}^X_{\ell'}$, with ${\rm X}=\{EB, (EE-BB)/2\}$.} reveals an excess for $\mathcal{D}^{EB}_\ell$, while it exhibits very good agreement for $\mathcal{D}^{(EE-BB)/2}_\ell$ across all CS methods. Specifically, the value of the $\chi^2_{\rm data}$ for $\mathcal{D}^{EB}_\ell$ ranges from 96 to 115 (depending on the CS method), whereas the expected $\chi^2$ value is 73, corresponding to the harmonic range $[42,1501]$. These values indicate a compatibility of the $EB$ spectrum with zero (expressed as the probability to exceed) ranging from $0.2\%$ for \texttt{Commander}, $0.3\%$ for \texttt{NILC}, $0.1\%$ for \texttt{SEVEM}, and $0.7\%$ for \texttt{SMICA}.

\section{Testing the harmonic-scale dependence of the isotropic cosmic birefringence}
\label{sec:test}

In this section, we first introduce the CB estimator from the CMB APS and the estimator we use to evaluate the CB at different angular scales. We then present our 
findings based on \textit{Planck} data.

\subsection{Estimators}

Starting from \cref{TTobs,EBobs}, it is possible to construct two harmonic-based estimators for $\beta$, known as $D$-estimators \cite{QUaD:2008ado, Gruppuso:2016nhj,Planck:2016soo}
\begin{align}
    D_\ell^{TB,\mathrm{obs}}(\hat \beta) &\equiv  C_\ell^{TB,\mathrm{obs}} \cos(2 \hat \beta) - C_\ell^{TE,\mathrm{obs}} \sin(2 \hat \beta) \, , \label{DTB} \\
     D_\ell^{EB,\mathrm{obs}} (\hat \beta)  &\equiv C_\ell^{EB,\mathrm{obs}} \cos(4 \hat \beta) - \frac{1}{2}\left(C_\ell^{EE,\mathrm{obs}}-C_\ell^{BB,\mathrm{obs}}\right) \sin(4 \hat \beta)\, , \label{DEB}
\end{align}
where $\hat \beta$ is the estimate of the CB angle $\beta$.
It can be shown, see e.g. \cite{Gruppuso:2016nhj}, that 
\begin{equation}
D_\ell^{Y,\mathrm{obs}}(\hat \beta = \beta) = 0 \,,
\label{mainproperty}
\end{equation}
for $Y=TB$ or $EB$. 
\Cref{mainproperty} suggests that $\beta$ can be found identifying the zeros of the $D$-estimators.
In the following, we base our analysis on $D_\ell^{EB,\mathrm{obs}}(\hat \beta)$, which is known to have greater constraining power than $D_\ell^{TB,\mathrm{obs}}(\hat \beta)$ for \emph{Planck}~\cite{Planck:2016soo}.\footnote{This statement is strictly true when the full available harmonic range is considered. It is possible that for very specific subranges the two estimators have closer statistical efficiency. We plan to return to this point in future investigations.}

\subsection{Estimates of the $\beta_{\ell}$}
\label{sec:estimatesbetaell}

\begin{figure}[!ht]
    \centering
    \begin{minipage}{0.49\textwidth}
        \centering
        \includegraphics[width=\textwidth]{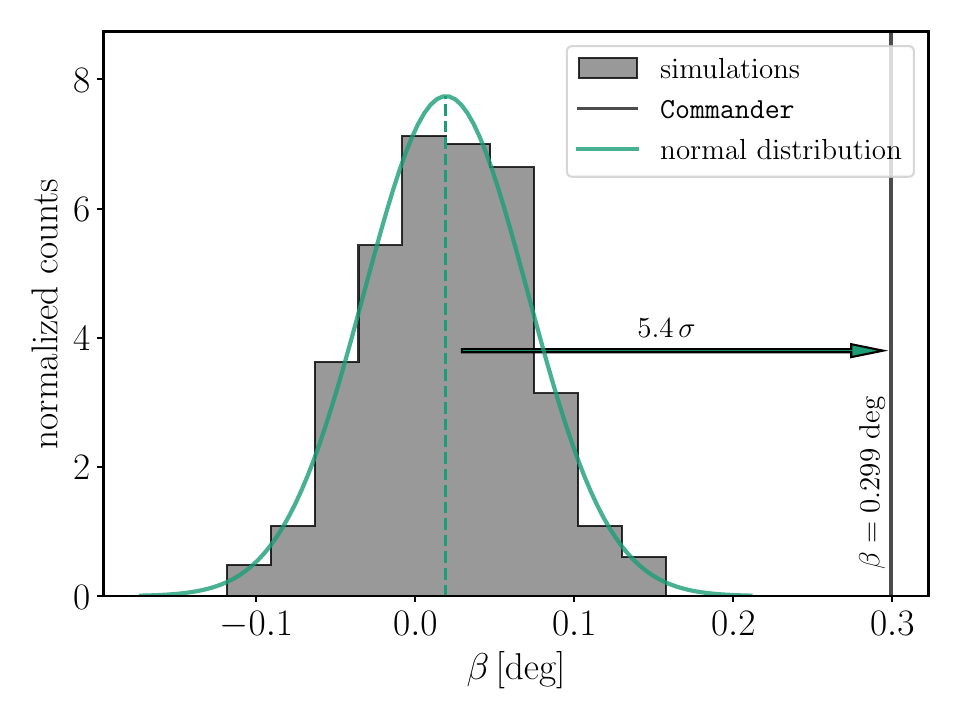}
    \end{minipage}
    \hfill
    \begin{minipage}{0.49\textwidth}
        \centering
        \includegraphics[width=\textwidth]{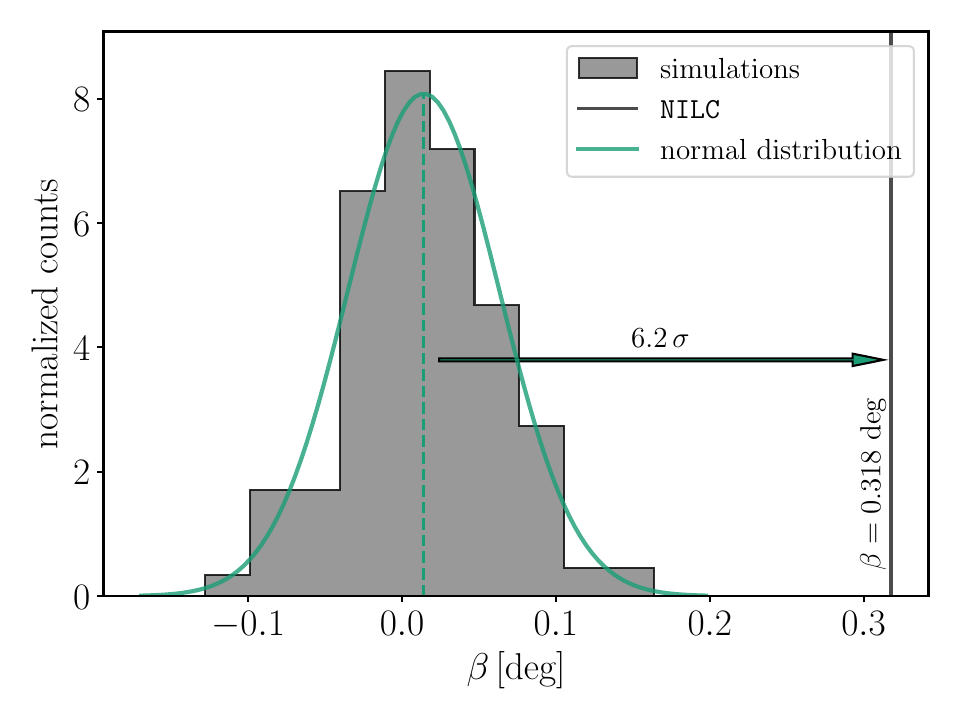}
    \end{minipage}
    \hfill
    \begin{minipage}{0.49\textwidth}
        \centering
        \includegraphics[width=\textwidth]{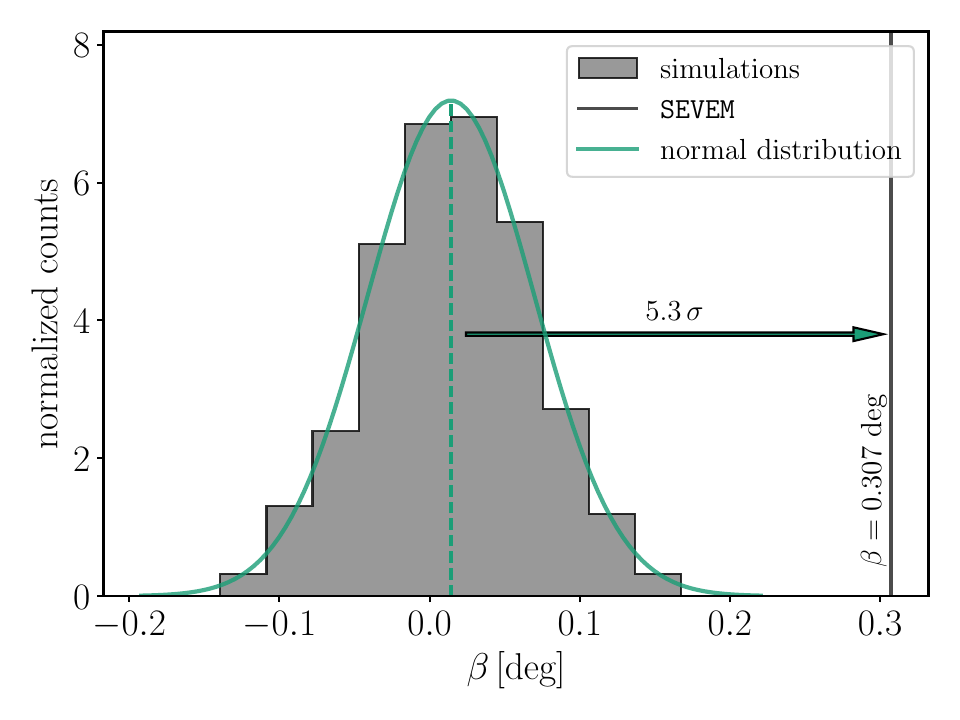}
    \end{minipage}
    \hfill
    \begin{minipage}{0.49\textwidth}
        \centering
        \includegraphics[width=\textwidth]{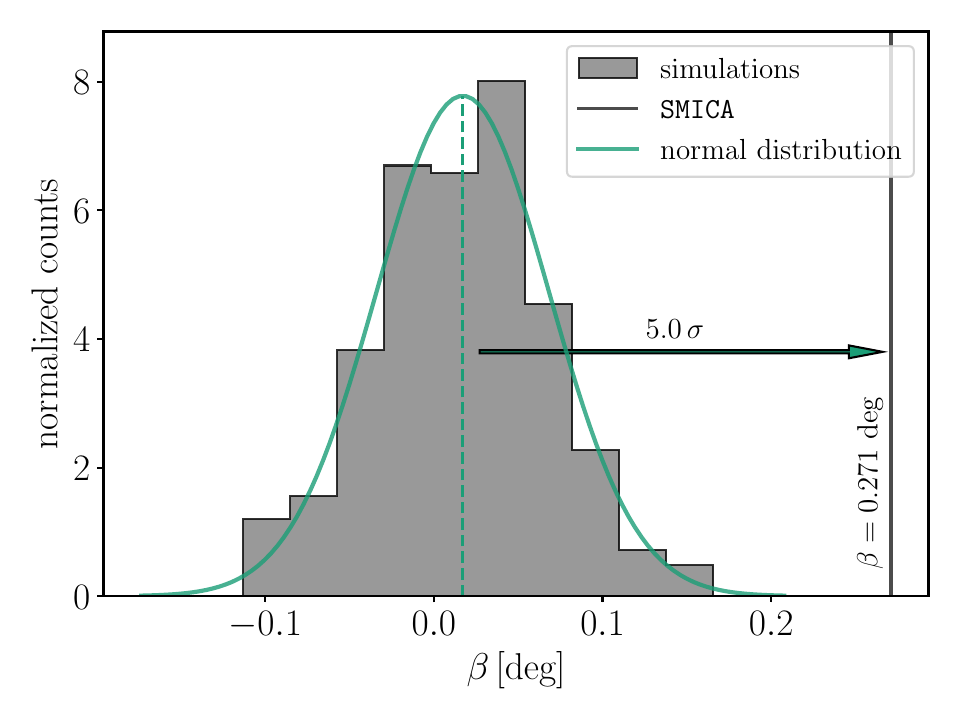}
    \end{minipage}
    \caption{Estimated $\beta$ for all the CS methods: upper left panel for \texttt{Commander}, upper right for \texttt{NILC}, lower left for \texttt{SEVEM} and lower right for \texttt{SMICA}. In each panel, the histogram (in gray), built with simulations, represents  the expected distribution for $\beta$, the vertical bar (in black) stands for the observed value. The curve (in green) is a Gaussian distribution, whose mean and standard deviation are derived from the corresponding histogram. This normal distribution is used to calculate the significance of the observed value (also reported above the arrow connecting the mean of the histograms to the estimated angle). }
    \label{fig:beta_whole_range}
\end{figure}

The first part of our analysis consists of recovering previous results already present in the literature. We estimate the CB angle by minimising the following $\chi^2(\hat \beta)$
\begin{equation}\label{chi2_EBb}
    \chi^2(\hat \beta)= \sum_{b,b'} D_{b}^{EB,\mathrm{obs}}(\hat \beta) (\mathrm{C}^{EB})^{-1}_{b b'} D_{b'}^{EB,\mathrm{obs}} (\hat \beta) \, ,
\end{equation}
considering the whole range of multipoles, where $b,b'=1,\dots,73$, corresponding to the harmonic range $[42,1501]$, which is equally spaced with $\Delta\ell =20$, and
with $\mathrm{C}^{EB}_{b b'}$ being the analytical covariance matrix for \textit{EB} computed through \texttt{NaMaster}, see \cref{angularpowerspectra}.
As also done in previous analyses, see e.g. \cite{Planck:2016soo}, we exclude the lowest multipoles to mitigate possible residual of systematic effects (including those of astrophysical origin). 
For each CS method, we compute $\beta$ for each simulation and for the corresponding \textit{Planck} CMB solution. These values were used to construct \cref{fig:beta_whole_range}, where the histograms represent the expected distribution from simulations for each CS method, and the vertical bars indicate the observed CB angle.
In each panel, we also display a Gaussian distribution, whose mean and standard deviation are derived from the corresponding histogram. This normal distribution is used to compute the significance of the observed value, which is also reported above the arrow connecting the mean of the histograms to the estimated angle. The probabilities associated with the observations range from $5 \sigma$ of the \texttt{SMICA} solution to $6.2 \sigma$ of \texttt{NILC}, whereas we find $5.4\sigma$ and $5.3\sigma$ for \texttt{Commander} and \texttt{SEVEM}, respectively.
The estimated means and standard deviations of the $\beta$s are reported in \cref{table:beta_WR}, along with the residual CB angle $\beta_{\rm res} = \langle \beta_i \rangle_{i\in \rm sims}$ computed on the simulations. 
Note that the latter corresponds to a fraction of the statistical uncertainty associated with the $\beta$ and is therefore not 
responsible for the overall measured CB effect. However, it can be interpreted as a source of systematic uncertainty.
\begin{table}[!ht]
\begin{tabular}{|l|c|c|}
\hline
\rowcolor[HTML]{EFEFEF} 
{\color[HTML]{000000} \textbf{Method}} & {\color[HTML]{000000} $\beta \text{\ [deg]}$} &  {\color[HTML]{000000} $\beta_{\text{res}} \text{\ [deg]}$} \\ \hline
PR3  \tt{Commander}            & $0.30 \pm 0.05$         &       $0.02$     \\ \hline
PR3 \tt{NILC}                  & $0.32 \pm 0.05$         &       $0.01$     \\ \hline
PR3 \tt{SEVEM}                 & $0.31 \pm 0.06$         &       $0.01$     \\ \hline
PR3 \tt{SMICA}                 & $0.27 \pm 0.05$         &       $0.02$     \\ \hline
\end{tabular}
\centering
\caption{Values of $\beta$ obtained over the harmonic range [42,1501] for the four CS methods considered in this work. In the third column, we report the mean value of the CB angle estimated from the simulations, $\beta_{\rm res}$.}
\label{table:beta_WR}
\end{table}

\begin{figure}[!ht]
    \centering
    \begin{minipage}{0.49\textwidth}
        \centering
        \includegraphics[width=\textwidth]{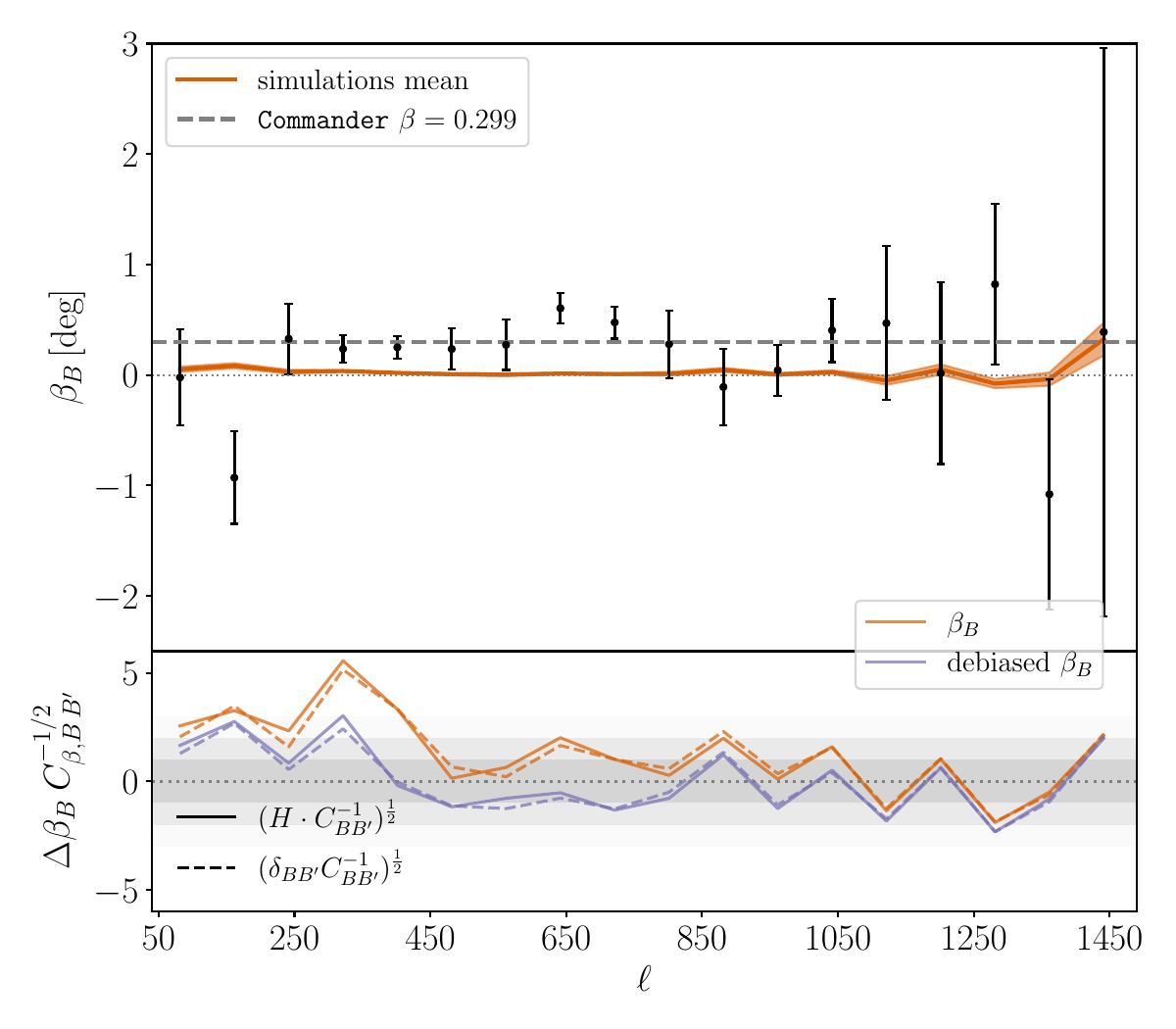}
    \end{minipage}
    \hfill
    \begin{minipage}{0.49\textwidth}
        \centering
        \includegraphics[width=\textwidth]{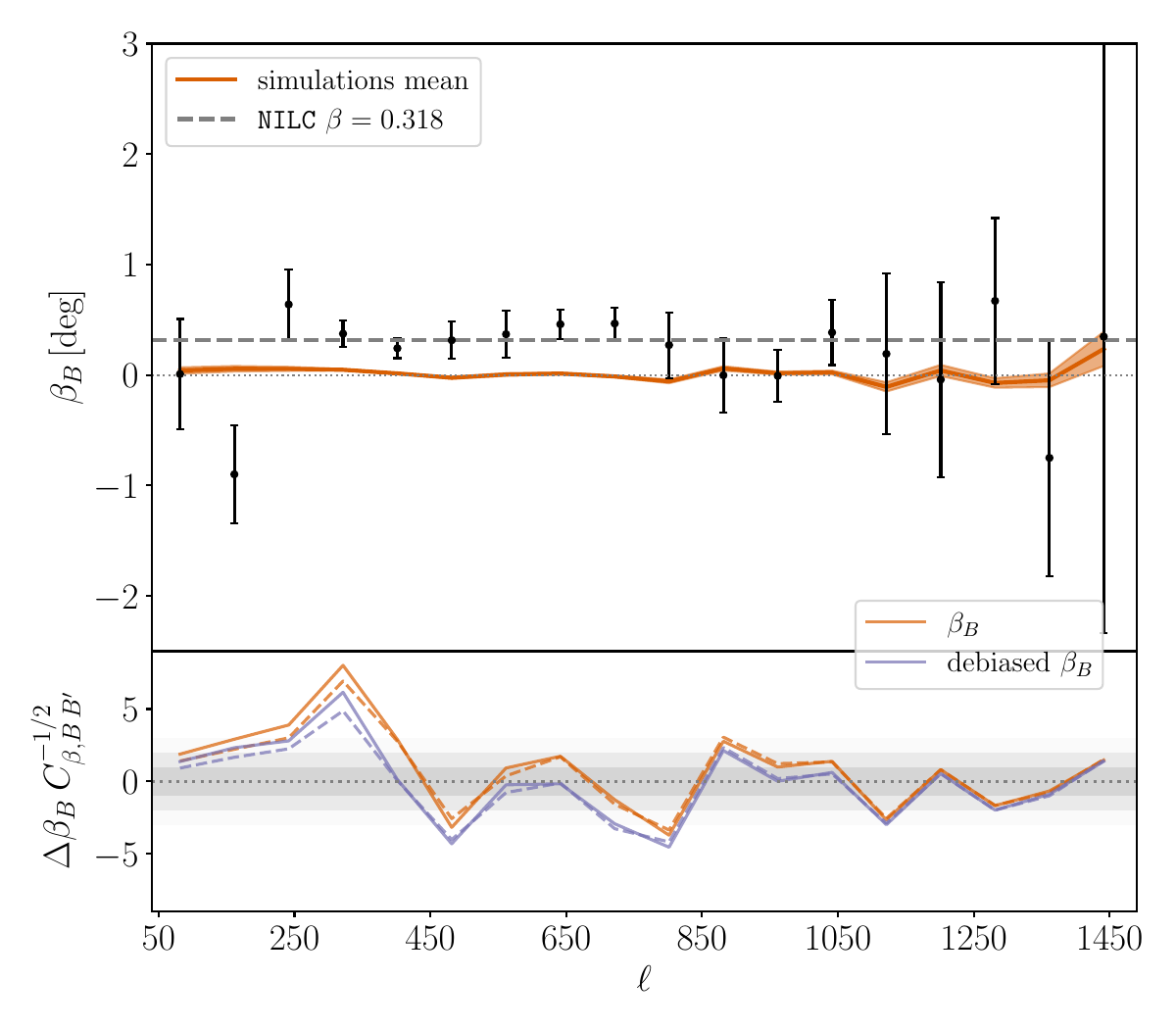}
    \end{minipage}
    \hfill
    \begin{minipage}{0.49\textwidth}
        \centering
        \includegraphics[width=\textwidth]{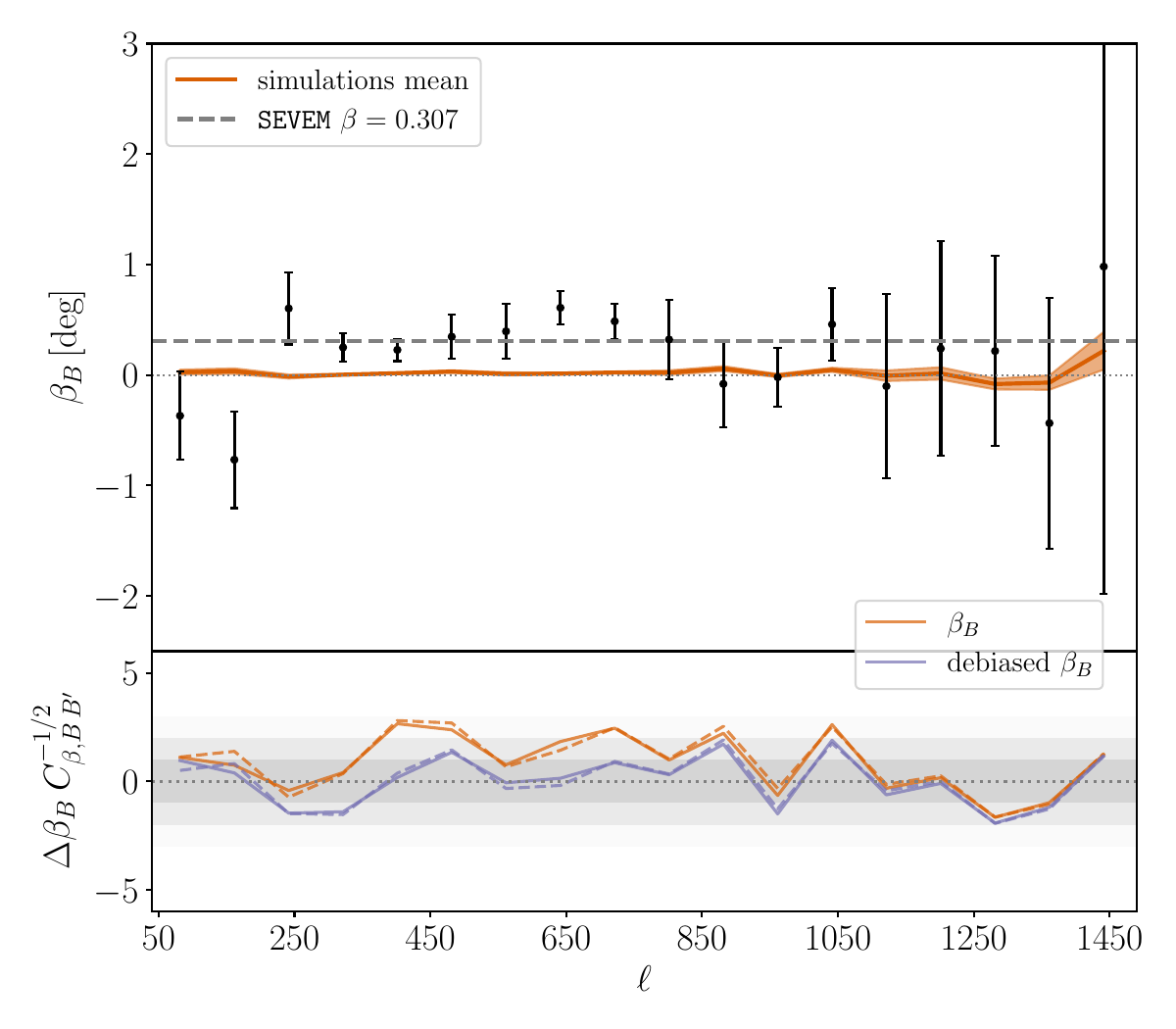}
    \end{minipage}
    \hfill
    \begin{minipage}{0.49\textwidth}
        \centering
        \includegraphics[width=\textwidth]{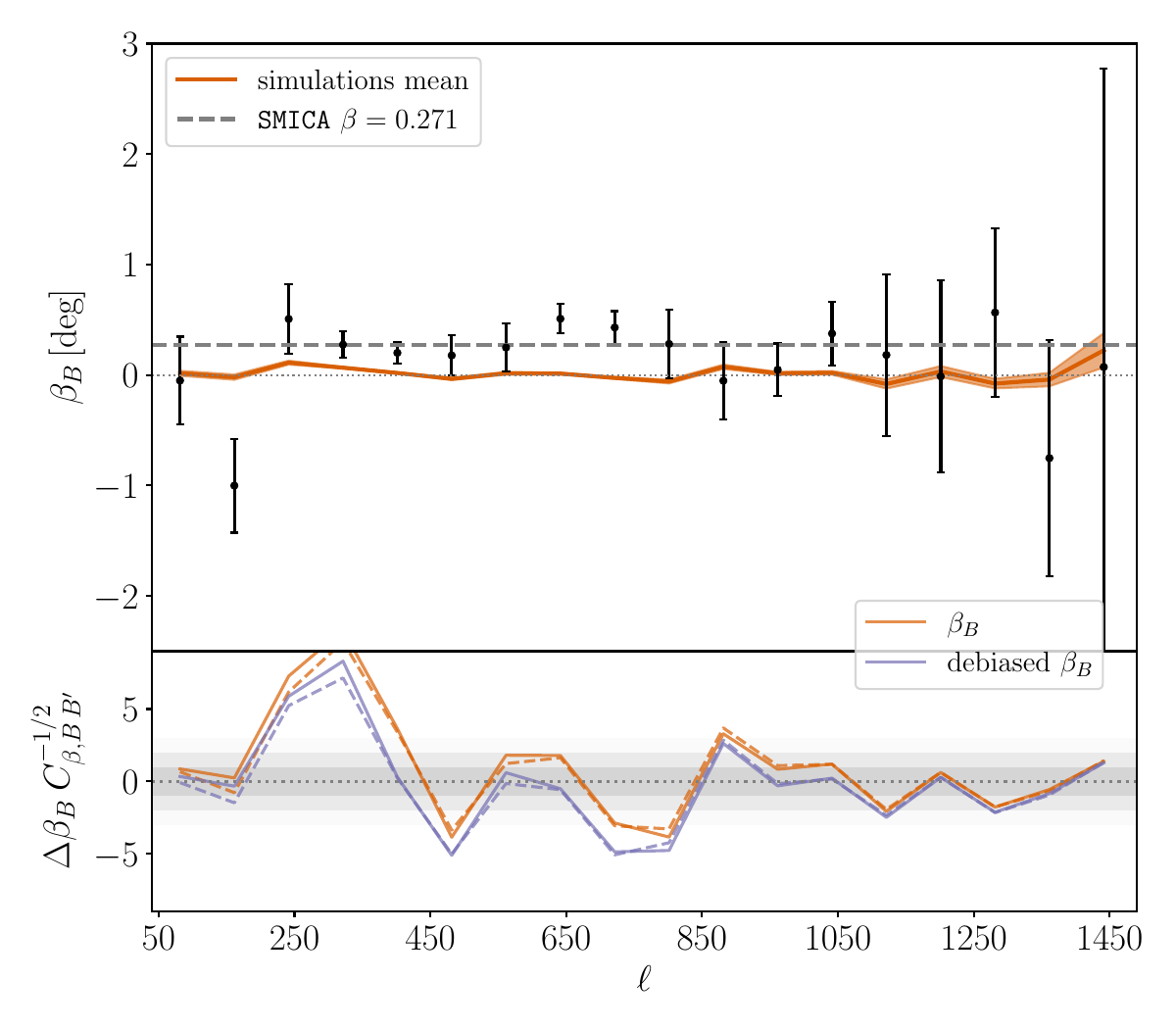}
    \end{minipage}
    \caption{Estimated $\beta_B$ for all the CS methods: upper left panel for \texttt{Commander}, upper right for \texttt{NILC}, lower left for \texttt{SEVEM}, and lower right for \texttt{SMICA}. In each panel we report the values of $\beta_B$ from data (black) with errorbars computed from the numerical covariance $\mathrm{C}_{B B'}^{\beta}\delta_{BB'}$ and the mean value, $\langle\beta_B\rangle$, with its $1\sigma$ area from the corresponding simulations (orange); grey dashed lines indicate the value of $\beta$ computed on the whole multipole range. In the lower tile we show the fluctuation of the simulations' mean with respect to the null hypothesis before removing the constant value of $\beta_{\rm res}$ at each multipole (orange) and after (indigo), and considering the full dense covariance matrix corrected by the Hartlap factor (solid) or only its diagonal (dashed). }
    \label{fig:EB_scatter_plot}
\end{figure}

We then evaluate the CB angle, $\beta_B$, over four harmonic bins, $b$, by minimizing the following $\chi_B^2(\hat \beta_B)$
\begin{equation}\label{chi2_EBB}
    \chi^2_{B}(\hat \beta_B)= \sum_{b,b^{\prime}} 
    D_{b}^{EB,\mathrm{obs}}(\hat \beta_B) (\mathrm{C}^{EB})^{-1}_{b b^{\prime}} D_{b^{\prime}}^{EB,\mathrm{obs}} (\hat \beta_B) \, ,
\end{equation}
where $B$ labels four aggregated consecutive bins $b$ in order to cover the multipole range [42,1481].\footnote{We excluded the last bin, $b=73$, to make the total number of bins divisible by four. 
Additionally, we have found that the estimate of $\beta$ on the aggregated bins corresponding to the multipole range $\ell\in[2,41]$, carries an uncertainty approximately 10 times larger than that obtained from the same analysis conducted on the aggregated harmonic bins in $\ell\in[42,1481]$. }
In this way $B \in [1,18]$ is mapping $b \in [1,72]$.
For each CS method and for each of the 300 corresponding simulations, we have computed the values $\beta_B$ that correspond to the minimum of \cref{chi2_EBB}.
We have then estimated their covariance from simulations as
\begin{equation}
    \mathrm{C}_{B B'}^{\beta} = \big\langle \left(\beta_B - \langle \beta_B\rangle \right)\left(\beta_{B'} -\langle \beta_{B'}\rangle\right) \big\rangle \, ,
\label{covariancebetab}
\end{equation}
where $\langle \beta_B\rangle$ is the mean over the simulations for each bin $B$. Using the same procedure, we estimate $\beta_B$ from \textit{Planck} data. 
Results are displayed in \cref{fig:EB_scatter_plot}. For each panel in the figure, we illustrate the compatibility of $\langle \beta_B\rangle$ with the null hypothesis in the lower tiles. This is shown through the quantity $\langle \beta_B\rangle \left[H \cdot (\mathrm{C}^{\beta}_{BB’})^{-1} \right]^{1/2}$ (solid orange lines), which is expected to lie mostly within $\pm 3\sigma$ for each bin $B$. The inverse covariance matrix is corrected using the Hartlap factor $H$ to account for the limited number of effective degrees of freedom in the simulation-based covariance estimate \cite{Hartlap:2006kj}. Note also that, as before, with the power of $1/2$ applied to a given matrix, we mean the Cholesky decomposition of the same matrix.
We also perform the same calculation using only the diagonal elements of $\mathrm{C}^{\beta}_{BB’}$, this time without applying the Hartlap correction (dashed lines). As can be seen in \cref{fig:EB_scatter_plot}, the off-diagonal terms contribute negligibly to the total uncertainty.
This test reveals only partial compatibility of the \textit{Planck} simulations processed with \texttt{Commander}, \texttt{NILC}, and \texttt{SMICA}, as evidenced by excess power around $\ell \approx 300$ for all three CS methods, and additionally around $\ell \approx 500$ \rm{and} $700$ for \texttt{NILC} and \texttt{SMICA}. To address this, we applied a debiasing procedure, subtracting the constant residual value $\beta_{\rm res}$ -- see \cref{table:beta_WR} -- from each $\beta_B$, separately for each CS method.
We then re-evaluated the standardised quantities and present the results -- now debiased -- in the same lower tiles of \cref{fig:EB_scatter_plot}, using indigo lines to indicate the corrected curves. This procedure restores full compatibility with the null hypothesis for \texttt{Commander}, whereas it only mildly mitigates the residual excess for \texttt{NILC} and \texttt{SMICA}. \texttt{SEVEM} exhibits full compatibility regardless of the debiasing procedure.
We will comment on the impact of the residual excess in the analysis section.
Nevertheless, we note that all estimates obtained from the different CS methods are in very good agreement, with relative fluctuations typically around $1 \sigma$ level.\footnote{This is not true in the excluded multipole range [2,41].}

\subsection{Fitting $\beta_{\ell}$ with a power-law like functional}
\label{sec:power-law}

In this section, we estimate the birefringence angles $\beta_B$ across angular scales using a two-parameter power-law model defined as
\begin{equation}
\beta(B) = \beta_0 \left(\frac{\ell_B}{\ell_{B_0}}\right)^n \,,
\label{eqn:power_law}
\end{equation}
where $\beta_0$ is the amplitude, $n$ is the power (or slope) that characterizes the harmonic scale dependence 
at a reference pivot scale $\ell_{B_0}$.
The power index $n$ is the key parameter in this analysis, as it quantifies deviations from a constant birefringence angle across multipoles.
Since the power-law form in \cref{eqn:power_law} is sensitive to the choice of pivot, we fix $B_0$ to the bin which minimises the correlation between $\beta_0$ and $n$, while also providing the smallest systematic uncertainty coming from $\beta_{\rm res}$. 
This corresponds to the sixth bin, with $\ell_{B_0=6} \in [442, 522]$, which is consistent across all four CS methods, see \cref{app:pivot_scale} for details.
We developed the following pipeline in order to test the power-law model in \cref{eqn:power_law}:
\begin{enumerate}
    \item we construct a two-dimensional grid $(\beta_0, n)$ centered at $(0,0)$. It consists of \(1000 \times 1000\) points, spanning the ranges $\beta_0 \in [-1,1] \ \text{deg}$ and $n \in [-1.5, 2.5]$;
    \item we then evaluate the value of a 2-dimensional $\chi^2$ at each value of the grid $(\beta_0, n)$, assuming the theoretical power law in \cref{eqn:power_law};
    \item we minimize the 2-dimensional $\chi^2$ to find the best-fit values $(\beta_0, n)$.
\end{enumerate}

The $\chi^2(\beta_0, n)$ has the following expression  
\begin{equation}\label{eqn:chi2EB}
    \chi^2(\beta_0, n) = \sum_{B,B'}^{18} \big[\beta(B) - \beta_B \big] \left(\mathrm{C}^{\beta}\right)^{-1}_{BB'} \big[\beta(B') - \beta_{B'} \big] \,,  
\end{equation}
where $\chi^2(\beta_0, n)$ quantifies the goodness of fit for each point on the grid.
This procedure has been applied to $\beta_B$ as obtained from each CS methods, to $\beta_{\rm res}$ to propagate the systematic effect to the parameters, and to the {\it debiased} dataset, $\beta_B - \beta_{\rm res}$.
\begin{table}[tb]
\centering
\renewcommand{\arraystretch}{1.3} 
\begin{tabular}{|c@{\hskip 6pt}|l@{\hskip 6pt}|c@{\hskip 6pt}|c@{\hskip 6pt}|c@{\hskip 6pt}|c@{\hskip 6pt}|}
\hline
\rowcolor[HTML]{EFEFEF} 
\textbf{Par.} & \textbf{Data} & \texttt{Commander} & \texttt{NILC} & \texttt{SEVEM} & \texttt{SMICA} \\ \hline

\multirow{3}{*}{$\beta_0\,\mathrm{[deg]}$} 
& $\beta_B$                    & $0.28\pm 0.05$   & $0.31\pm 0.05$   & $0.30\pm 0.06$   & $0.26\pm 0.05$   \\ 
& $\beta_{\rm res}$            & $0.019\pm 0.003$ & $0.013\pm 0.003$ & $0.013\pm 0.003$ & $0.017\pm 0.003$ \\ 
& $\beta_B - \beta_{\rm res}$  & $0.26\pm 0.05$   & $0.29\pm 0.05$   & $0.29\pm 0.06$   & $0.25\pm 0.05$   \\ \hline

\multirow{3}{*}{$n$} 
& $\beta_B$                    & $0.60^{+0.34}_{-0.31}$ & $0.29^{+0.31}_{-0.29}$ & $0.57^{+0.33}_{-0.31}$ & $0.52^{+0.36}_{-0.34}$ \\ 
& $\beta_{\rm res}$            & $0.00^{+0.35}_{-0.31}$ & $0.00^{+0.45}_{-0.41}$ & $0.00^{+0.49}_{-0.41}$ & $0.00^{+0.38}_{-0.33}$ \\ 
& $\beta_B - \beta_{\rm res}$  & $0.62^{+0.36}_{-0.33}$ & $0.30\pm 0.31$ & $0.59^{+0.34}_{-0.32}$ & $0.54^{+0.38}_{-0.35}$ \\ \hline
\end{tabular}
\caption{Best-fit values and $1\sigma$ errors for $(\beta_0, n)$ from different CS methods, and different inputs: data $\beta_B$, bias $\beta_{\rm res}$, and debiased data $\beta_B - \beta_{\rm res}$.}
\label{table:best_fit_couples}
\end{table}

\begin{figure}[h!]
    \centering
    \begin{subfigure}[b]{0.49\textwidth}
        \centering
        \includegraphics[width=\textwidth]{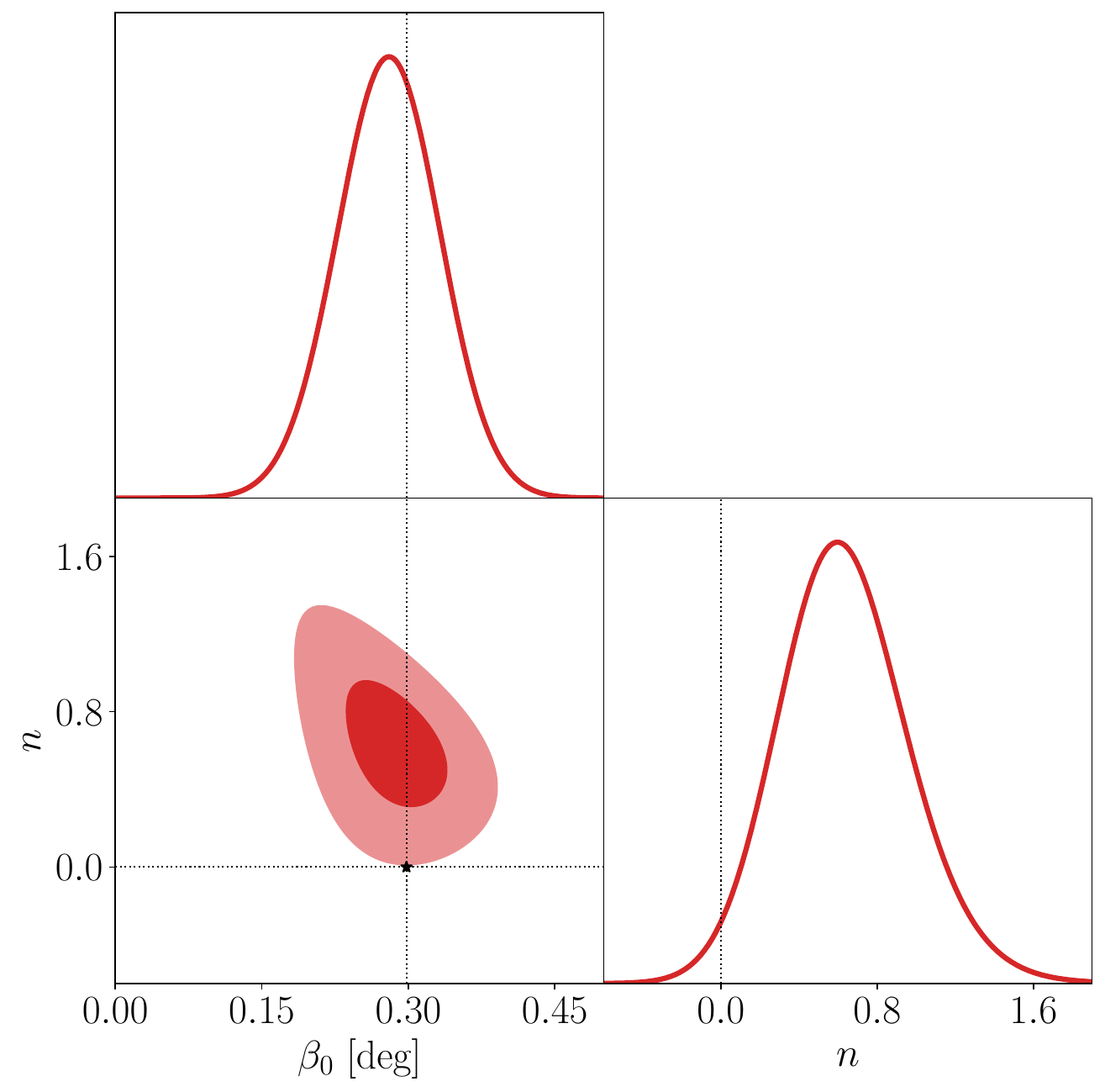}
        \label{fig:commander_contplotEB}
    \end{subfigure}
    \hfill
    \begin{subfigure}[b]{0.49\textwidth}
        \centering
        \includegraphics[width=\textwidth]{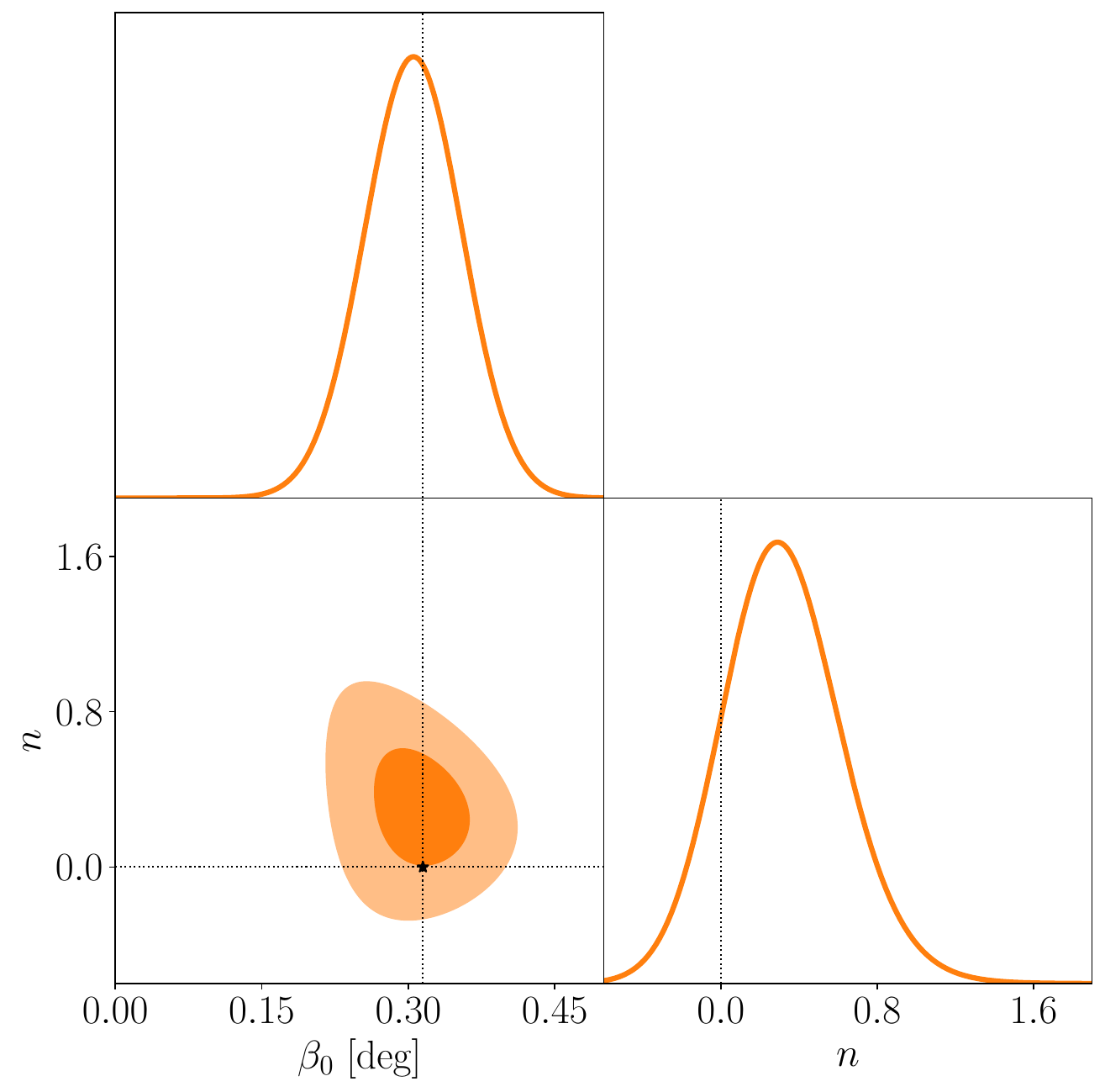}
        \label{fig:nilc_contplotEB}
    \end{subfigure}
    
    \vspace{0.5cm}
    \begin{subfigure}[b]{0.49\textwidth}
        \centering
        \includegraphics[width=\textwidth]{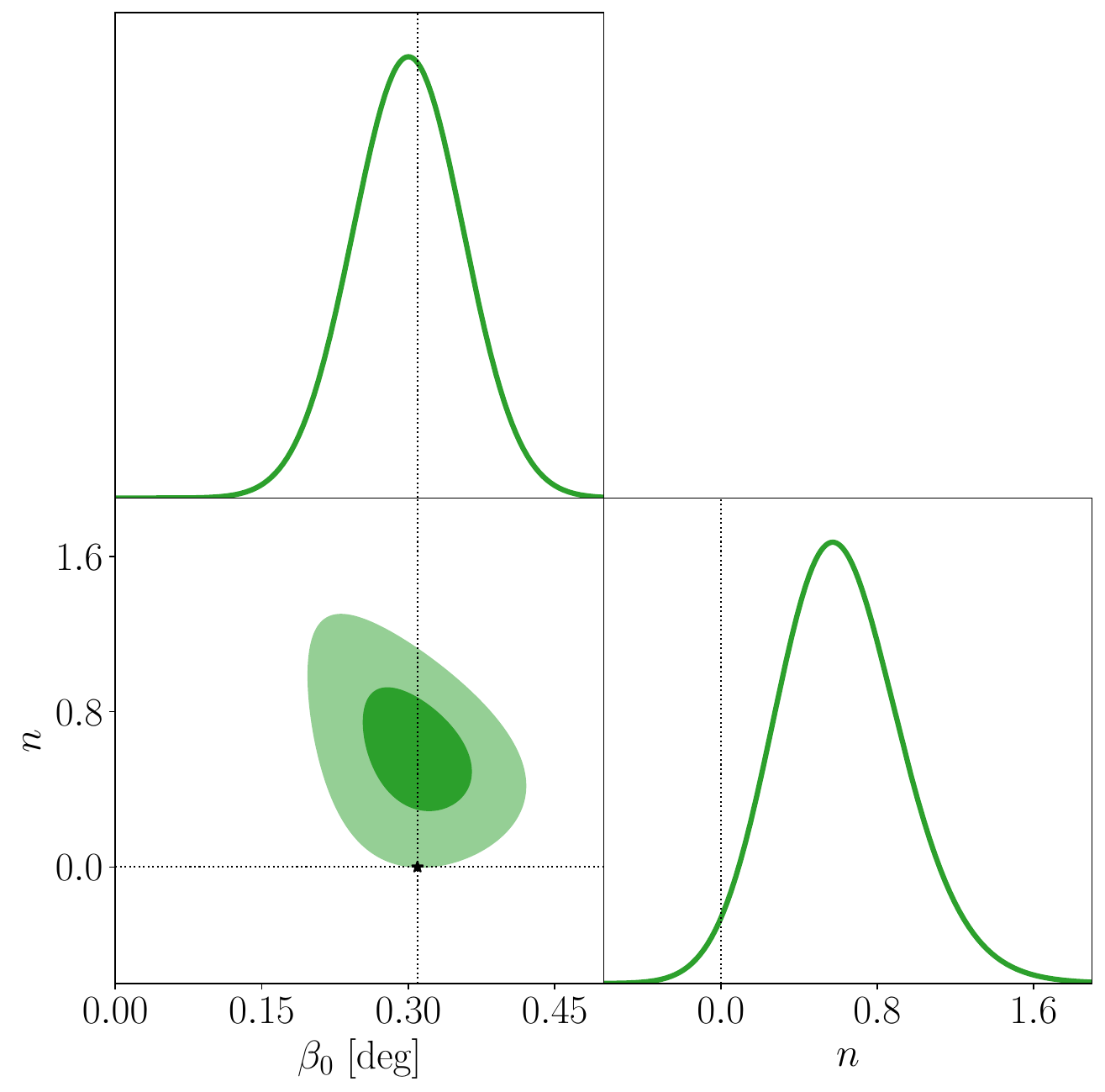}
        \label{fig:sevem_contplotEB}
    \end{subfigure}
    \hfill
    \begin{subfigure}[b]{0.49\textwidth}
        \centering
        \includegraphics[width=\textwidth]{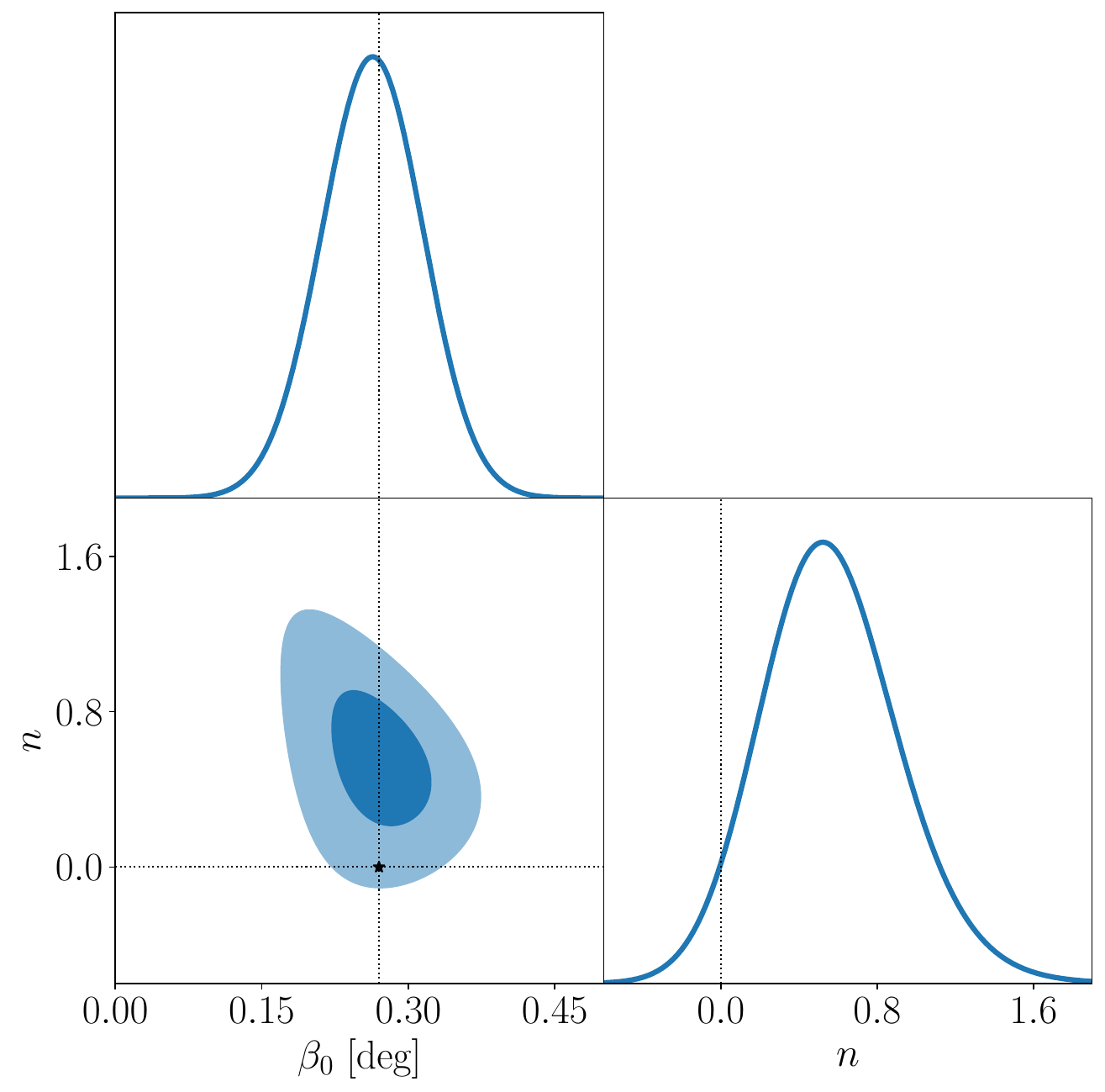}

        \label{fig:smica_contplotEB}
    \end{subfigure}
    
    \caption{Triangular plots illustrating the power-law constraints for all four CS methods using $D^{EB}$ estimator: upper left panel for \texttt{Commander}, upper right for \texttt{NILC}, lower left for \texttt{SEVEM}, and lower right for \texttt{SMICA}. The $\star$ represent $(\beta,0)$ with $\beta$ given in \cref{table:beta_WR}.}
    \label{fig:contourplots_EB}
\end{figure}

We display the marginal distributions for $(\beta_0, n)$ in \cref{fig:contourplots_EB}, and report in \cref{table:best_fit_couples} the constraints on $\beta_0$ and $n$.
The constraints on the input $\beta_{\rm res}$ represent a systematic effect that contributes only to $\beta_0$ at the level of $\sim 40\%$ of the statistical uncertainty. In particular, the systematic uncertainty on $n$ peaks at zero and therefore possible residuals from the \textit{Planck} analysis pipeline do not represent a significant limitation to our analysis. 
Our estimates of $n$ are compatible with zero within the total error budget at the level of $1.8\sigma$ for \texttt{Commander}, $1.0\sigma$ for \texttt{NILC}, $1.7\sigma$ for \texttt{SEVEM}, and $1.5\sigma$ for \texttt{SMICA}. 
However, when we build templates for \texttt{NILC} and \texttt{SMICA} in order to cope with the excess found at $\ell \approx 300$, $500$ and $700$, see again \cref{fig:EB_scatter_plot}, we obtain
$\beta_0 = 0.30 \pm 0.05 \, \mathrm{[deg]}$ and $n=0.40 \pm 0.33$ for \texttt{NILC}
and 
$\beta_0 = 0.25 \pm 0.05 \, \mathrm{[deg]}$ and $n=0.68 \pm 0.36$ for \texttt{SMICA}.
This means that the shift in the power index, $n$, is $-0.11$ for \texttt{NILC} and $-0.16$ for \texttt{SMICA}, corresponding to fractions of a standard deviation of $0.36\sigma$ and $0.51\sigma$, respectively. This suggests a non-negligible interplay between foreground contamination and systematic residuals; accounting for this yields updated estimates of the power index that are compatible among the different CS methodologies within one $\sigma$.
The templates are built using the means of $\beta_B$ computed from simulations (without debiasing, see the orange lines in \cref{fig:EB_scatter_plot}), selecting only those fluctuations above $3.5 \sigma$. We then estimate $\beta_0$ and $n$ on the dataset after subtracting the template. 

\subsection{Fitting $\beta_{\ell}$ with a non-parametric Bayesian reconstruction}
\label{sec:Nonparametric}
To model the data in a way that avoids theoretical biases, we adopt a model-independent reconstruction approach based on a flexible representation using moving nodes or knots. This method provides a versatile framework that allows the data to guide the shape of the reconstructed function without relying on a specific parametric model, as we did in the previous section.

Reconstruction is performed by placing a set of nodes along the domain of the data. The amplitudes and positions of these nodes are treated as free parameters to be inferred from the data performing a Bayesian analysis. The function between the nodes is interpolated using a linear spline. This approach, commonly known as \texttt{FlexKnot} model \cite{Millea:2018bko}, allows the reconstruction to flexibly adapt to the underlying data structure.
To ensure the robustness and generality of the reconstruction, the two endpoints of the domain are fixed at $\ell_\mathrm{min} = 42$ and $\ell_\mathrm{max} = 1481$, while the intermediate nodes are allowed to vary in a sorted manner, i.e. ordered such that $\ell_i < \ell_{i+1}$ for any node $i$ to avoid encountering in a non-physical form.
The number of nodes is increased incrementally, starting with one node ($N=1$), corresponding to a constant amplitude, and progressively adding more nodes ($N=2,3,4,\dots$). This iterative approach allows us to assess the trade-off between model complexity and quality of fit. 

Inference is performed using the nested sampling algorithm implemented in \texttt{PyPolyChord} \cite{Handley:2015fda,Handley:2015vkr}\footnote{\url{https://github.com/PolyChord/PolyChordLite}} with 800 live points. 
We consider the isotropic birefringence angle estimates as obtained in \cref{sec:estimatesbetaell} for all the four CS methods. 
The log-likelihood function, i.e.  \cref{eqn:chi2EB}, incorporates the observed data and their uncertainties, while the priors are chosen to reflect reasonable physical constraints on the node amplitudes and positions. Amplitudes have been sampled with a uniform prior between $[-20^\circ,20^\circ]$ and positions have been sampled with a uniform prior between $[\ell_\mathrm{min},\ell_\mathrm{max}]$ 
applying a \textit{forced identifiability prior} transformation \cite{Handley:2015vkr,Handley:2019fll}.
For each configuration of the number of nodes, the posterior distributions of the parameters are obtained and the model evidence is computed. The Bayesian evidence provides a quantitative metric for comparing reconstructions with different numbers of nodes, allowing us to identify the optimal balance between model simplicity and explanatory power. 
The posterior distributions of the reconstructed functions are visualized using the \texttt{fgivenx} package \cite{fgivenx},\footnote{\url{https://github.com/handley-lab/fgivenx}} which provides a clear representation of the uncertainty in the reconstruction. 

\begin{figure}[h!]
    \centering
    \includegraphics[width=0.495\textwidth]{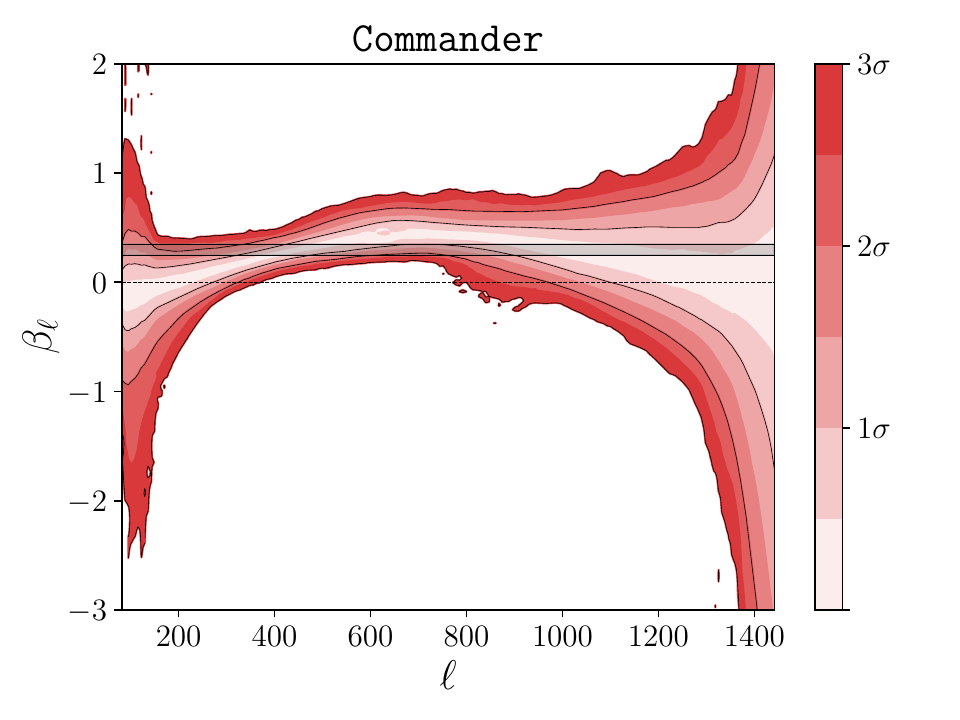}
    \includegraphics[width=0.495\textwidth]{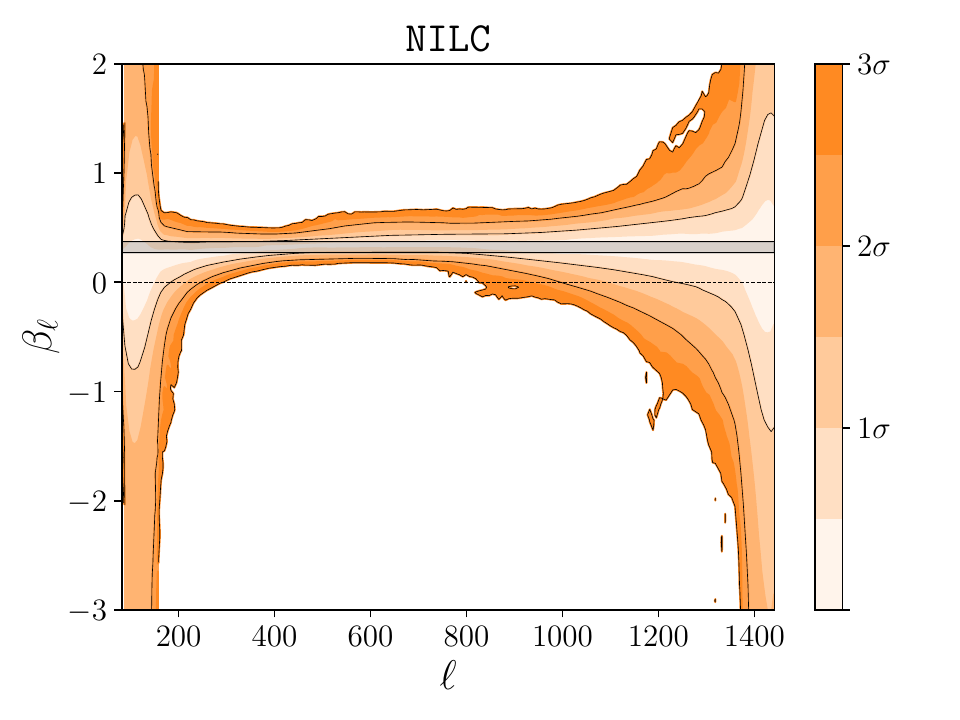}
    \includegraphics[width=0.495\textwidth]{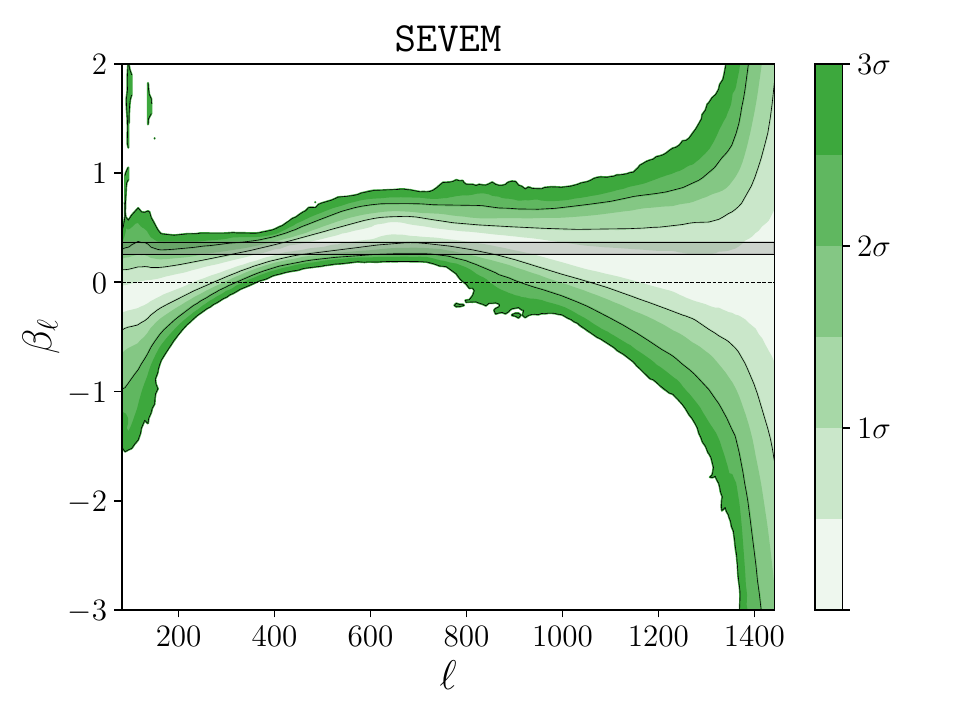}
    \includegraphics[width=0.495\textwidth]{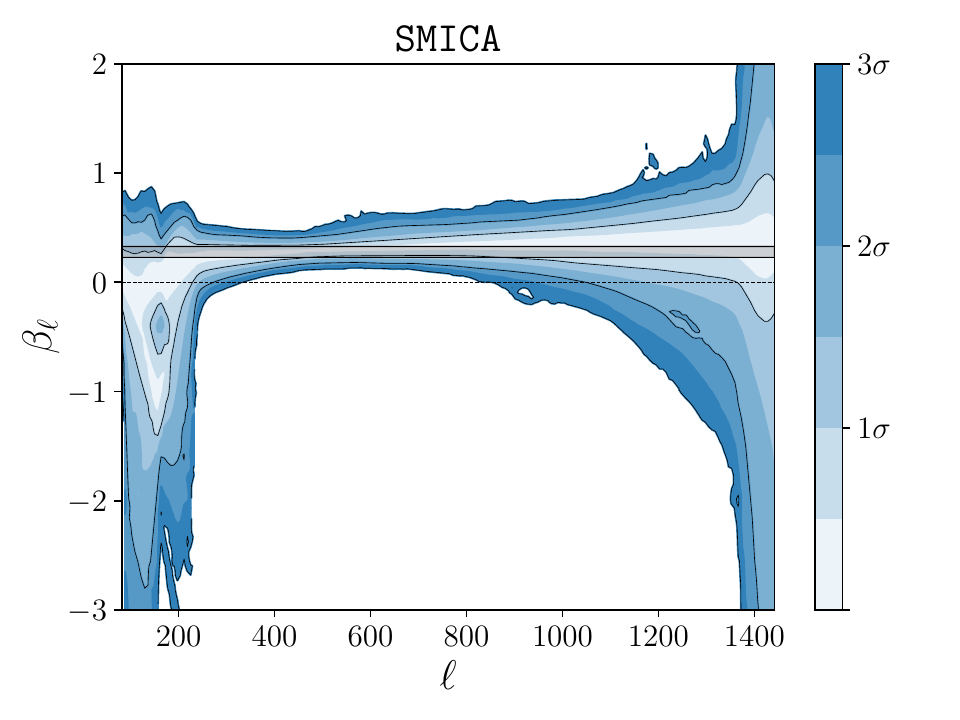}
    \caption{Reconstruction of the angular isotropic cosmic birefringence angle plotted using iso-probability credibility intervals in the $(\beta_\ell, \ell)$ plane, with their masses converted to $\sigma$ values via an inverse error function transformation, obtained from the $EB$-based estimator applied to the \texttt{Commander}, \texttt{NILC}, \texttt{SEVEM}, and \texttt{SMICA} maps, assuming four nodes ($N=4$). The grey band correspond to the result obtained with $N=1$ collected on \cref{tab:beta_N1}.
    The grey shadow regions correspond to the 68\% CL obtained performing the same analysis with one node ($N=1$).}
    \label{fig:reconstruction_N4}
\end{figure}
\Cref{fig:reconstruction_N4} shows the reconstructed function obtained with the four CS methods, assuming $N=4$. The shaded regions represent the credible intervals derived from the posterior samples. We compare the reconstructed $\beta_\ell$ spectra for $N=4$ with the results obtained in the single-node case, i.e. $N=1$, displayed as grey shadow region in \cref{fig:reconstruction_N4}, and reported in \cref{tab:beta_N1}.
While all four CS methods yield consistent results, broadly agreeing with a constant CB angle, small deviations at 68\% CL appear at large angular scales ($\ell \lesssim 350$) for \texttt{Commander}, \texttt{SEVEM}, and \texttt{SMICA}, where the results are more compatible with a vanishing angle.
\begin{table}[h!]
\begin{tabular}{|l|l|}
\hline
\rowcolor[HTML]{EFEFEF} 
{\color[HTML]{000000} \textbf{Method}} & {\color[HTML]{000000} $\beta \text{\ [deg]}$} \\ \hline
PR3  \tt{Commander}            & $0.30 \pm 0.05$                                 \\ \hline
PR3 \tt{NILC}                  & $0.32 \pm 0.05$                                 \\ \hline
PR3 \tt{SEVEM}                 & $0.31 \pm 0.06$                                 \\ \hline
PR3 \tt{SMICA}                 & $0.28 \pm 0.05$                                 \\ \hline
\end{tabular}
\centering
\caption{Mean values of $\beta$ and uncertainties at 68\% CL obtained with \texttt{PolyChord} considering one node ($N=1$).}
\label{tab:beta_N1}
\end{table}

In \cref{app:Nonparametric_EB}, we collect the results obtained from this pipeline varying the number of nodes compared to the results showed here for $N=4$ as well as the Bayes factor calculated with respect to the case with one node.

\section{Forecasts for future CMB experiments}
\label{sec:forecasts}

In this section, we present forecasts on our ability to recover $\beta_0$ and $n$ using upcoming CMB experiments. In particular, we focus on the following experiments: LiteBIRD (LB) \cite{LiteBird:2022,delaHoz:2025uae}, Simons Observatory - Large Aperture Telescope (SO-LAT) \cite{Ade_2019,SimonsObservatory:2025wwn}, and CMB Stage 4 (CMB-S4) \cite{Abazajian_2022,CMBs4:2024}.

For each experiment, we produce 500 simulated datasets, with the same underlying cosmological parameters, including a scale-invariant angle $\beta$ of $0.3 \rm \ deg$, but with different noise phases with the experimental configurations as reported in the upper block of \cref{tab:forecast-setup}.
We then apply the analysis pipeline presented in \cref{sec:power-law} to constrain $\beta_0$ and $n$. We use the same pivot scale as in the \textit{Planck} analysis to make the estimate of $n$ from the forecasts to be comparable with the one from the analysis in this work.\footnote{This choice does not necessarily represent the optimal choice for different experimental set-ups.} The forecasts for $\beta_0$ and $n$ are presented in \cref{table:forecast_beta}. The uncertainty on $\beta_0$ is up to $5$ times lower than that obtained with \textit{Planck} data, and the error on $n$ can be up to $\sim 7$ times lower than the larger error found with \textit{Planck} data.

\begin{table}[tb]
\centering
    \begin{tabular}{|l|l|l|l|}
    \hline
    \rowcolor[HTML]{EFEFEF} 
    {\color[HTML]{000000} \textbf{Parameter}} & {\color[HTML]{000000} LiteBIRD} & {\color[HTML]{000000} SO-LAT} & {\color[HTML]{000000} CMB-S4} \\ \hline
    beam $[\rm arcmin]$                  & $30$   & $1.4$  & $1$    \\ \hline
    sensitivity $[\mu K\cdot\rm arcmin]$ & $3.26$ & $8.49$ & $4.24$ \\ \hline
    sky fraction                         & $70\%$ & $40\%$ & $2\%$  \\ \hline
    \hline                              
    $N_{\rm side}$                      & $1024$ & $1024$ & $2048$ \\ \hline
    $\ell_{\rm min}$                      &  $11$ & $11$ & $51$  \\ \hline
    $\ell_{\rm max}$                        & $811$  & $3011$ & $3011$ \\ \hline
    \end{tabular}
    \caption{Summary of the experimental specifications for the considered upcoming CMB surveys (upper block), and the set-up for the analysis pipeline (lower block).}
    \label{tab:forecast-setup}
\end{table}

\begin{table}[h!]
\centering
\renewcommand{\arraystretch}{1.3} 
\begin{tabular}{|c@{\hskip 6pt}|c@{\hskip 6pt}|c@{\hskip 6pt}|c@{\hskip 6pt}|}
\hline
\rowcolor[HTML]{EFEFEF} 
\textbf{Par.} & LiteBIRD & SO-LAT & CMB-S4 \\ \hline

$\beta_0$\,[deg] & $0.30\pm 0.01$ & $0.30\pm 0.01$ & $0.30\pm 0.02$ \\ \hline
$n$       & $0.00\pm 0.07$ & $0.00\pm 0.05$ & $0.00\pm 0.07$ \\ \hline
\end{tabular}

\caption{Forecasted best-fit values and $1\sigma$ errors for $(\beta_0, n)$ from upcoming CMB experiments.}
\label{table:forecast_beta}
\end{table}

\section{Conclusions}
\label{sec:conclusions}

In this work, we
use the \textit{Planck} PR3 data and simulations to constrain the dependence of the cosmic birefringence angle $\beta$ on the harmonic scale $\ell$. 
Such a dependence might stem from ChS models if the mass of the pseudo-scalar field is not sufficiently small. 
In this case, the main observational signature is represented by the $EB$ spectrum that is not a perfect rescaling of $(EE-BB)/2$, a condition that would break the degeneracy between the CB isotropic angle and the instrumental polarisation angle.

Using the data and simulations as described in \cref{sec:descriptiondata}, we confirm the presence of an excess of $EB$ signal across all four CS methods, whose origin remains unclear. We find that the estimated $\beta$ over the whole harmonic range, i.e. [42,1501], is consistent with the literature \cite{Minami:2020odp, Diego-Palazuelos:2022dsq} regardless of the impact of the instrumental polarisation angle.

We have studied the harmonic dependence of this excess, through two different methods: a parametric approach, where we fit the data using the power law in \cref{eqn:power_law}, and a non-parametric approach based on the \texttt{FlexKnot} model.
The excess is found to be independent from the harmonic scale, at the $\lesssim 2\sigma$ level, for all CS methods. 
Using the non-parametric approach with the \texttt{FlexKnot} model 
we find a strong preference for a constant $\beta$ angle across the harmonic range (see \cref{fig:reconstruction_Bayes}). 

The main outcome of this work ultimately lies in the \textit{Planck} data preference for a constant value of the CB angle across the harmonic scales \cite{Eskilt:2023nxm,Namikawa:2025sft,Nakatsuka:2022epj}.
Future CMB experiments will further reduce \textit{Planck} uncertainties up to a factor of $7$. If the effect is due to new physics, these findings would provide strong constraints on the ultra-light nature of the axion field with even better perspectives for future probes.

\appendix
\section{Pivot scale} \label{app:pivot_scale}

\begin{figure}[h!]
    \centering
    \begin{subfigure}[b]{0.49\textwidth}
        \centering
        \includegraphics[width=\textwidth]{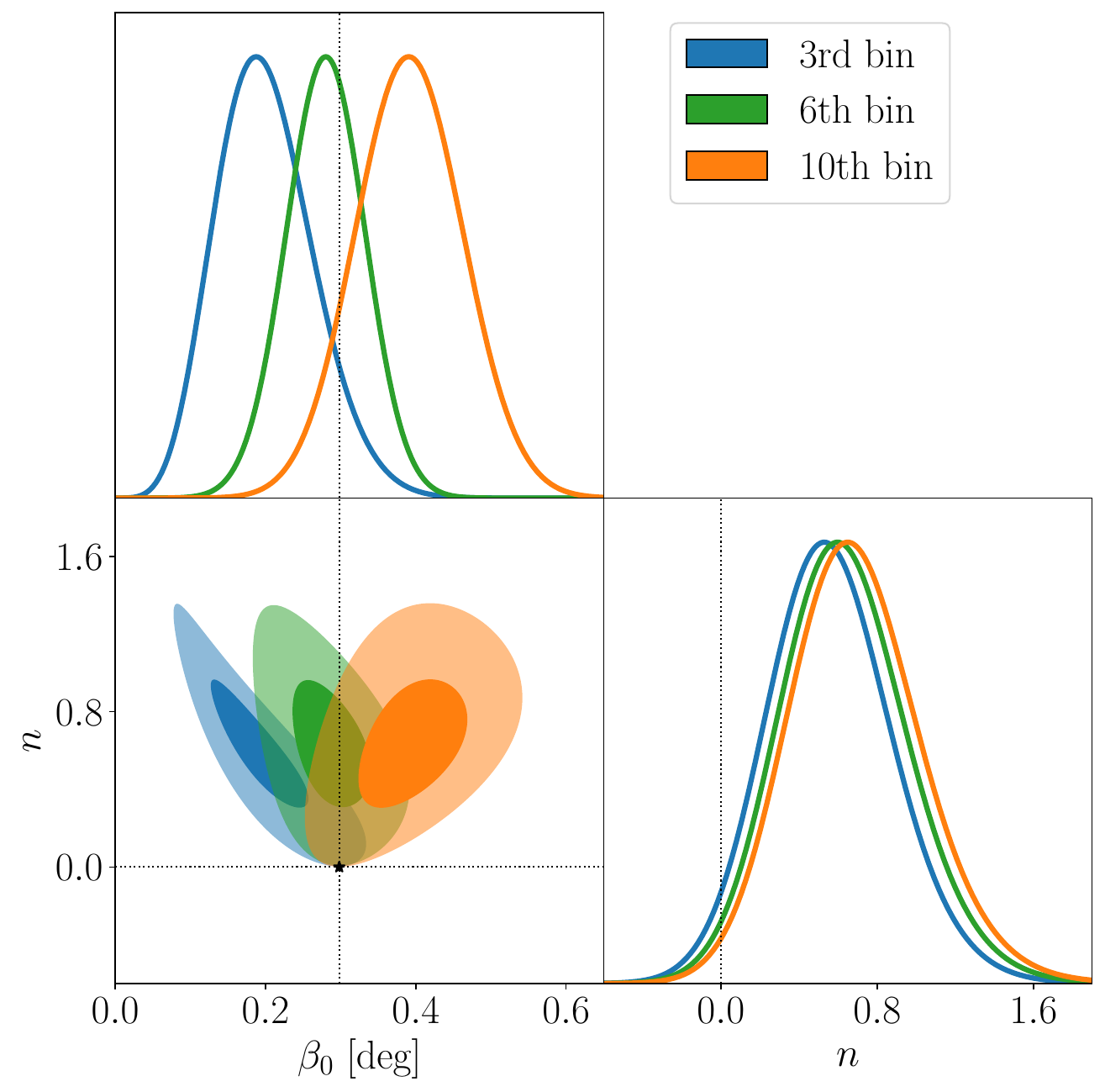}
        \label{fig:commander_contplotEB_pivot}
    \end{subfigure}
    \hfill
    \begin{subfigure}[b]{0.49\textwidth}
        \centering
        \includegraphics[width=\textwidth]{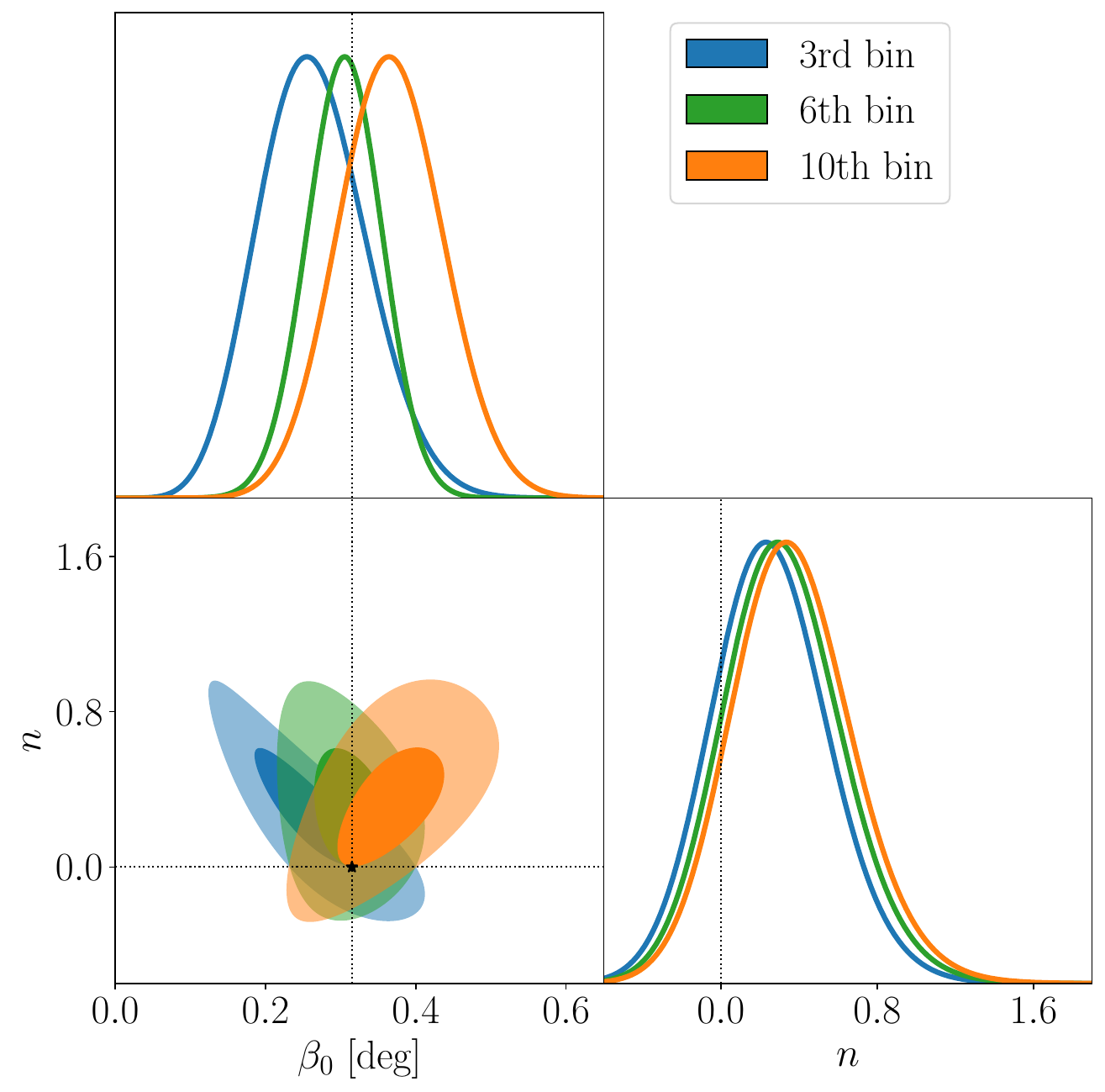}
        \label{fig:nilc_contplotEB_pivot}
    \end{subfigure}
    
    \vspace{0.5cm}
    \begin{subfigure}[b]{0.49\textwidth}
        \centering
        \includegraphics[width=\textwidth]{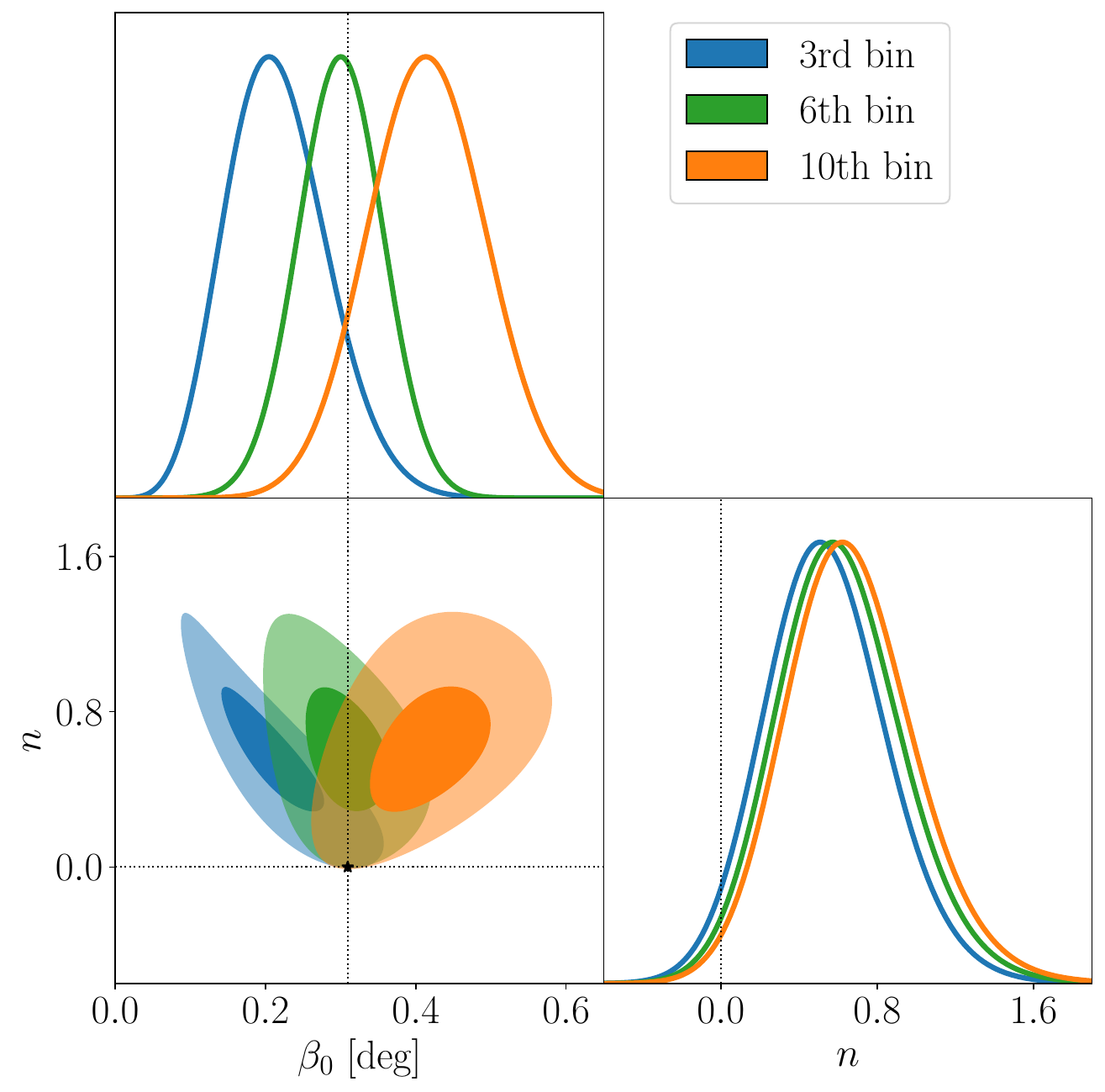}
        \label{fig:sevem_contplotEB_pivot}
    \end{subfigure}
    \hfill
    \begin{subfigure}[b]{0.49\textwidth}
        \centering
        \includegraphics[width=\textwidth]{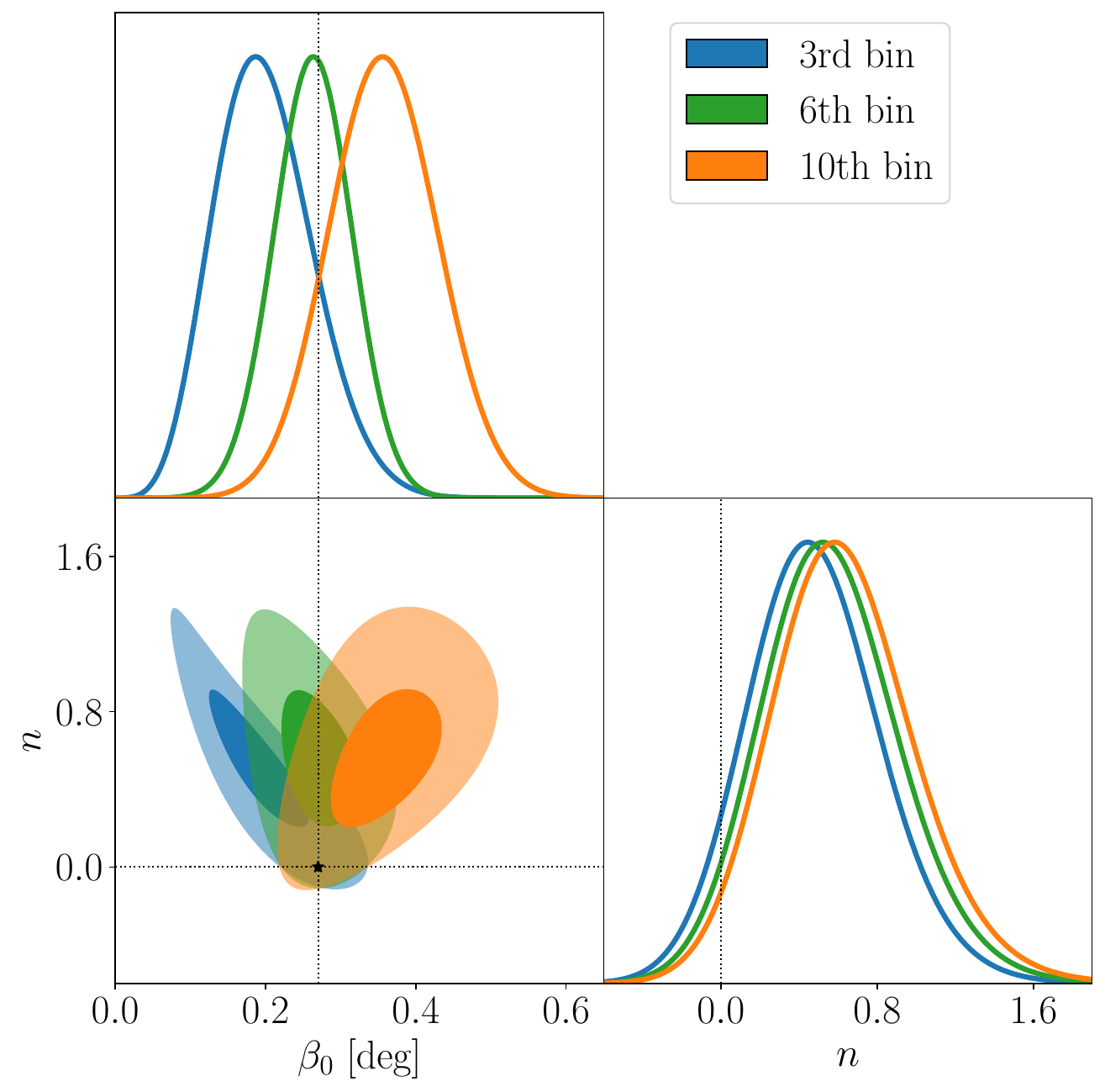}
        \label{fig:smica_contplotEB_pivot}
    \end{subfigure}
    \caption{Triangular plots reporting the power-law test results for all four CS methods: \texttt{Commander} (top left), \texttt{NILC} (top right), \texttt{NILC} (bottom left) and \texttt{SMICA} (bottom right), using $D^{EB}$ estimator and varying the pivotal bin (3rd, 6th or 10th).}
    \label{fig:contourplots_EB_bin_3_6_10}
\end{figure}

We assess the choice of the pivot scale $\ell_{B_0}$ by 
examining three different choices for $B_0$ across the four CS \textit{Planck} datasets. We select the bin that results in the smallest correlation between $\beta_0$ and $n$ while being compatible with the value of $\beta$ estimated 
across the entire range and yielding the smallest possible systematic uncertainty on $n$ from $\beta_{\rm res}$. \Cref{table:beta_B_bin_3_6_10,table:beta_res_bin_3_6_10} report our numerical results for the four CS datasets and for three choices of $B_0$ corresponding to the third ($\ell\in[202,281]$), sixth ($\ell\in[442,521]$) and tenth ($\ell\in[762,841]$) bins. We also show in \cref{fig:contourplots_EB_bin_3_6_10} the corresponding contour plots. We find that a common value of $B_0=6$ minimises both the correlation between the model parameters and the systematic uncertainty for $n$. 
This position corresponds to the region most constrained by the non-parametric reconstruction; see \cref{fig:reconstruction_N4}.

\begin{table}[h!]
\centering
\renewcommand{\arraystretch}{1.3}

\begin{tabular}{|c@{\hskip 6pt}|l@{\hskip 6pt}|c@{\hskip 6pt}|c@{\hskip 6pt}|c@{\hskip 6pt}|c@{\hskip 6pt}|}
\hline
\multirow{2}{*}{\textbf{Par.}} & \multirow{2}{*}{\textbf{Pivot scale}} & \multicolumn{4}{c|}{\cellcolor[HTML]{EFEFEF}\textbf{ Data considered: $\boldsymbol{\beta_B}$}} \\ \cline{3-6}
& & \texttt{Commander} & \texttt{NILC} & \texttt{SEVEM} & \texttt{SMICA} \\ \hline

\multirow{3}{*}{$\beta_0$\,[deg]} 
& 3rd bin       & $0.19^{+0.07}_{-0.06}$ & $0.26\pm 0.07$ & $0.20^{+0.07}_{-0.06}$ & $0.19^{+0.07}_{-0.06}$ \\ 
& 6th bin       & $0.28\pm 0.05$         & $0.31\pm 0.05$ & $0.30\pm 0.06$         & $0.26\pm 0.05$         \\ 
& 10th bin      & $0.39\pm 0.07$         & $0.36\pm 0.07$ & $0.41\pm 0.08$         & $0.36\pm 0.07$         \\ \hline

\multirow{3}{*}{$n$} 
& 3rd bin       & $0.53^{+0.33}_{-0.31}$ & $0.23^{+0.30}_{-0.29}$ & $0.51^{+0.32}_{-0.30}$ & $0.44^{+0.35}_{-0.33}$ \\ 
& 6th bin       & $0.60^{+0.34}_{-0.31}$ & $0.29^{+0.31}_{-0.29}$ & $0.57^{+0.33}_{-0.31}$ & $0.52^{+0.36}_{-0.34}$ \\ 
& 10th bin      & $0.65^{+0.35}_{-0.32}$ & $0.34^{+0.32}_{-0.30}$ & $0.62^{+0.34}_{-0.31}$ & $0.58^{+0.37}_{-0.34}$ \\ \hline
\end{tabular}

\caption{Best-fit values and $1\sigma$ errors for $(\beta_0, n)$ at different pivot scales (3rd, 6th, and 10th bin corresponding respectively to $\ell_{B_0} = 242, 482, 802$) for the various CS methods.}
\label{table:beta_B_bin_3_6_10}
\end{table}

\begin{table}[h!]
\centering
\renewcommand{\arraystretch}{1.3}

\begin{tabular}{|c@{\hskip 6pt}|l@{\hskip 6pt}|c@{\hskip 6pt}|c@{\hskip 6pt}|c@{\hskip 6pt}|c@{\hskip 6pt}|}
\hline
\multirow{2}{*}{\textbf{Par.}} & \multirow{2}{*}{\textbf{Pivot scale}} & \multicolumn{4}{c|}{\cellcolor[HTML]{EFEFEF}\textbf{Data considered: $\boldsymbol{\beta_{\rm res}}$}} \\ \cline{3-6}
& & \texttt{Commander} & \texttt{NILC} & \texttt{SEVEM} & \texttt{SMICA} \\ \hline

\multirow{3}{*}{$\beta_0$\,[deg]} 
& 3rd bin       & $0.019\pm 0.005$ & $0.014\pm 0.005$ & $0.014\pm 0.005$           & $0.017\pm 0.005$ \\ 
& 6th bin       & $0.019\pm 0.003$ & $0.013\pm 0.003$ & $0.013\pm 0.003$           & $0.017\pm 0.003$ \\ 
& 10th bin      & $0.019\pm 0.004$ & $0.014\pm 0.004$ & $0.014^{+0.005}_{-0.004}$ & $0.017\pm 0.004$ \\ \hline

\multirow{3}{*}{$n$} 
& 3rd bin       & $-0.07^{+0.33}_{-0.29}$ & $-0.12^{+0.42}_{-0.38}$ & $-0.12^{+0.45}_{-0.38}$ & $-0.08^{+0.36}_{-0.31}$ \\ 
& \textbf{6th bin}       & $\bf 0.00^{+0.35}_{-0.31}$  & $\bf 0.00^{+0.45}_{-0.41}$  & $\bf 0.00^{+0.49}_{-0.41}$  & $\bf 0.00^{+0.38}_{-0.33}$  \\ 
& 10th bin      & $0.05^{+0.36}_{-0.32}$  & $0.09^{+0.47}_{-0.42}$  & $0.10^{+0.53}_{-0.44}$  & $0.06^{+0.39}_{-0.34}$  \\ \hline
\end{tabular}

\caption{Best-fit values and $1\sigma$ errors for $(\beta_0, n)$ at different pivot scales (3rd, 6th, and 10th bin), estimated from the bias term $\beta_{\rm res}$ for various CS methods. We highlighted with bold text the sixth bin row corresponding to the minimal value recovered for $n$.}
\label{table:beta_res_bin_3_6_10}
\end{table}

\section{Non-parametric Bayesian reconstruction for $N \ne 4$ nodes} \label{app:Nonparametric_EB}
\begin{figure}[h!]
    \centering
    \includegraphics[width=0.45\textwidth]{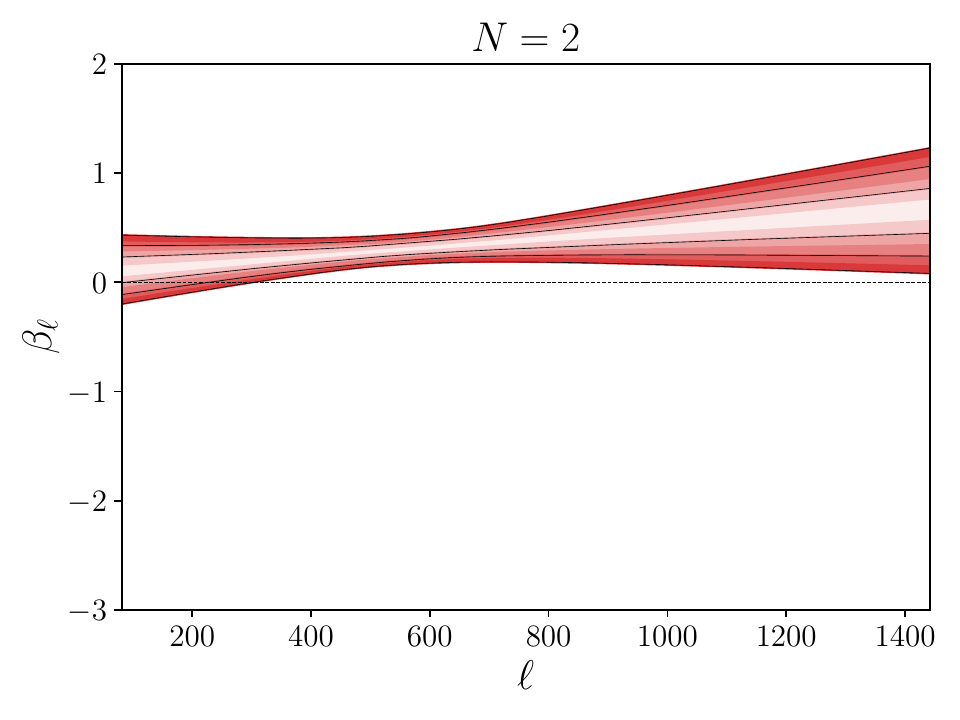}
    \includegraphics[width=0.45\textwidth]{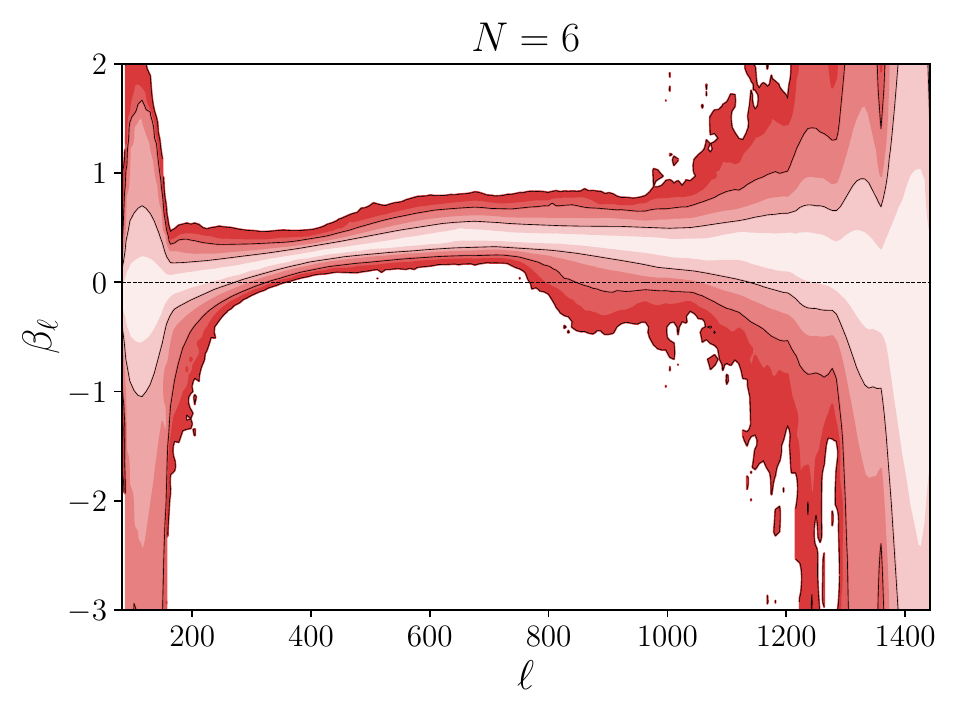}
    \includegraphics[width=0.45\textwidth]{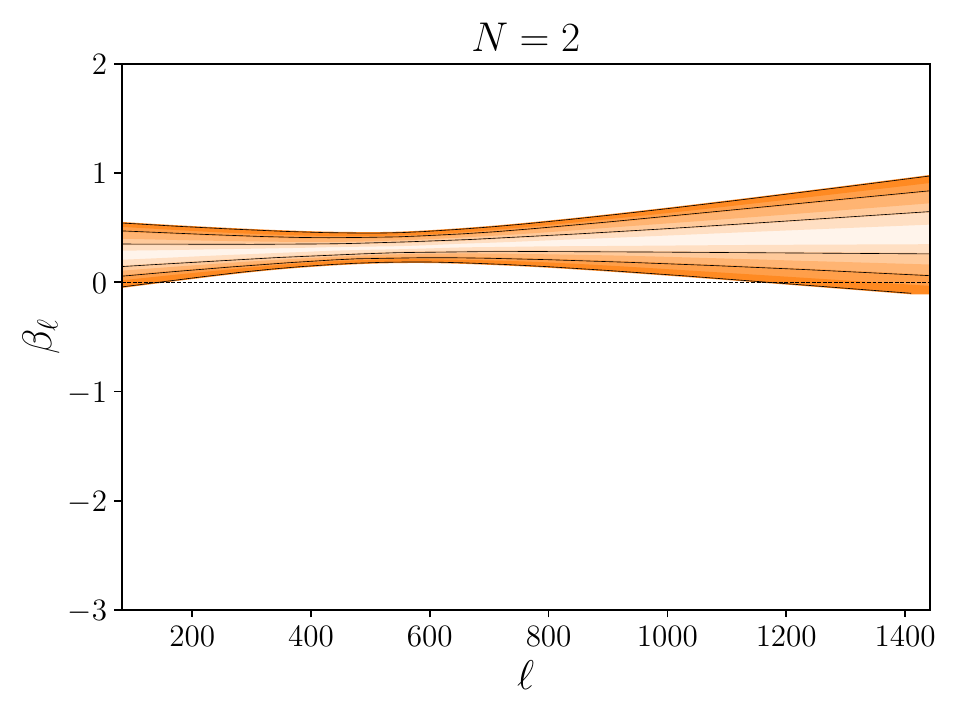}
    \includegraphics[width=0.45\textwidth]{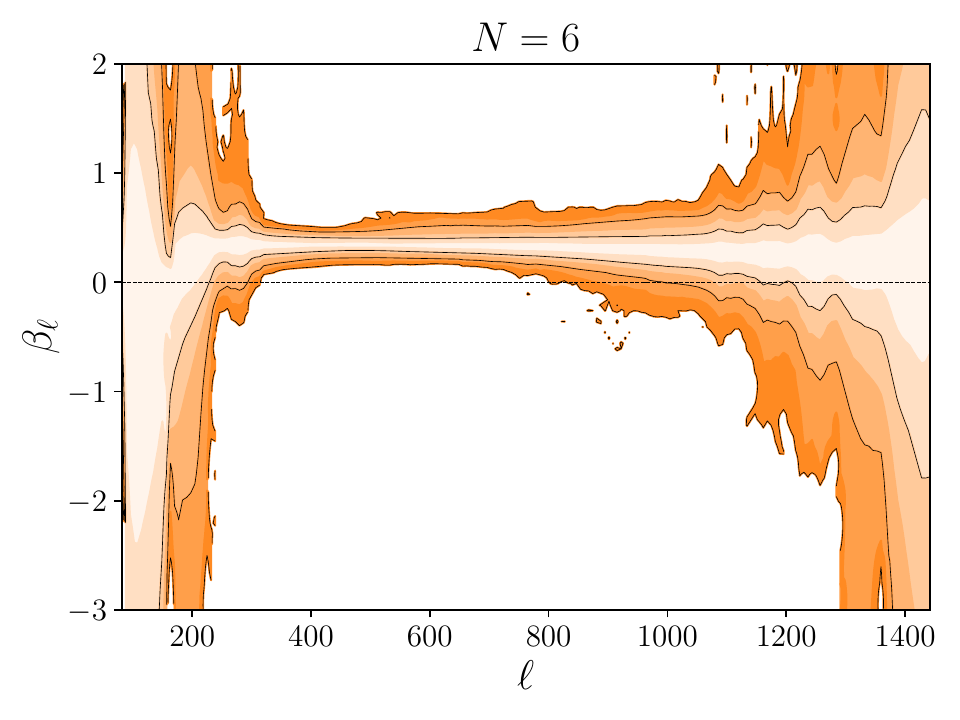}
    \includegraphics[width=0.45\textwidth]{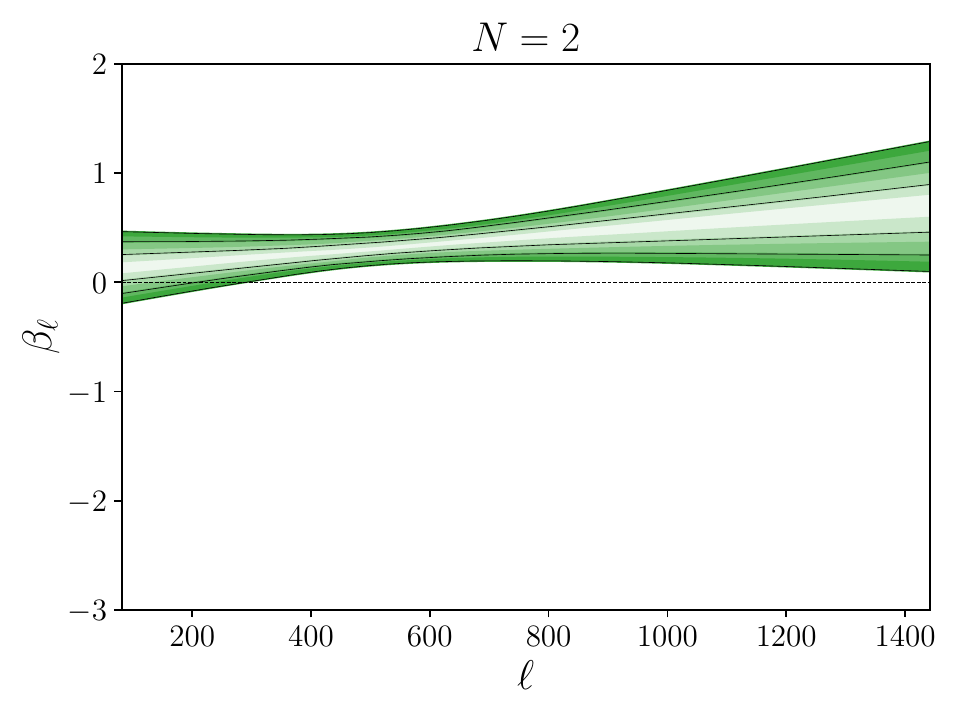}
    \includegraphics[width=0.45\textwidth]{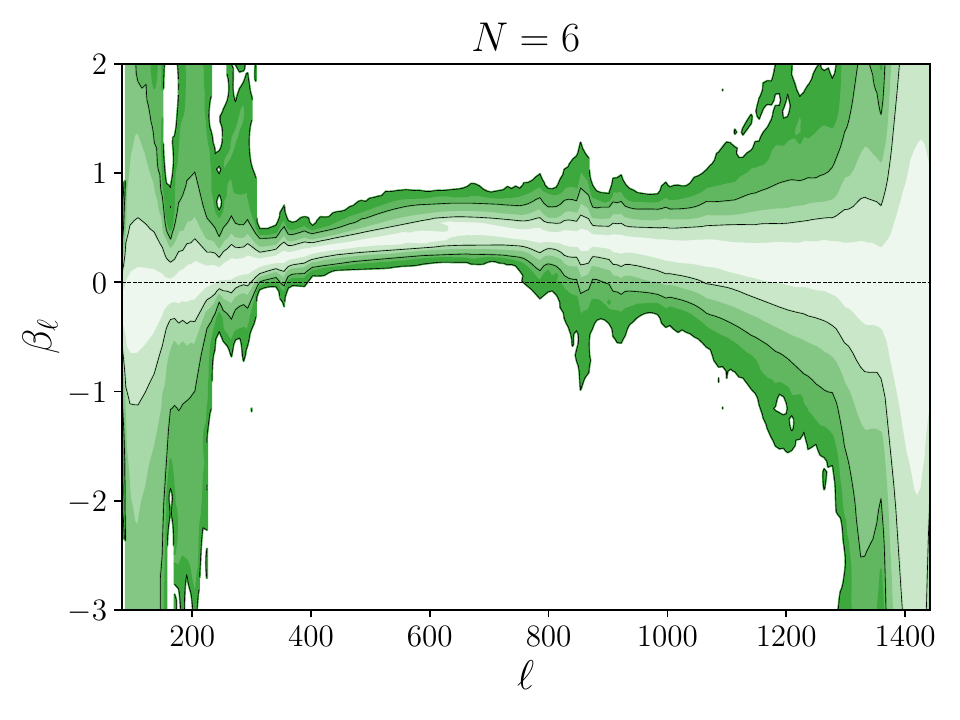}
    \includegraphics[width=0.45\textwidth]{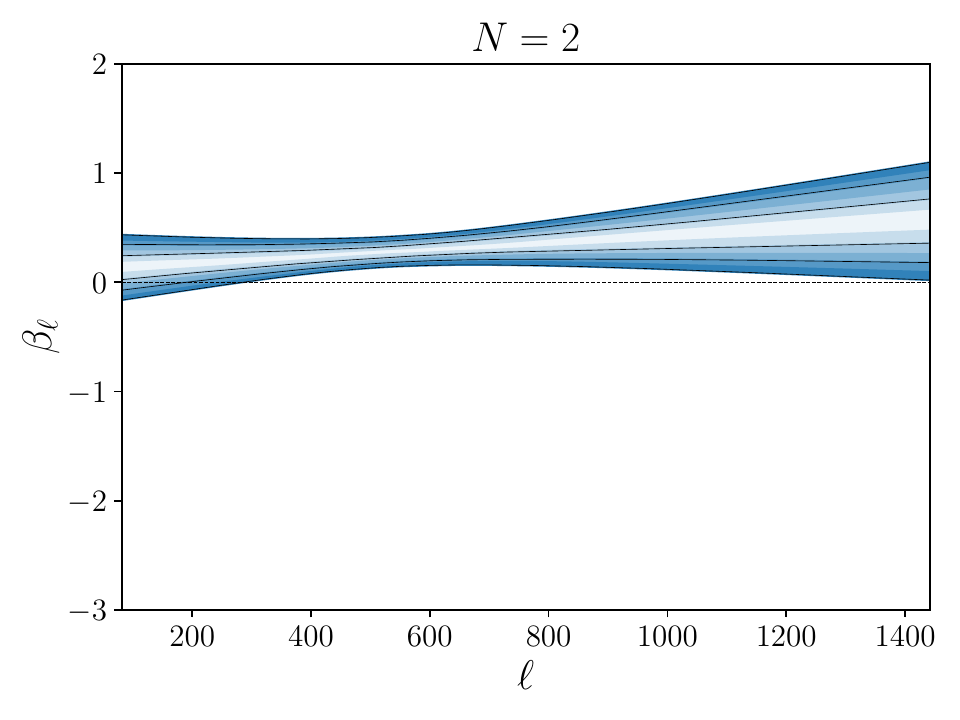}
    \includegraphics[width=0.45\textwidth]{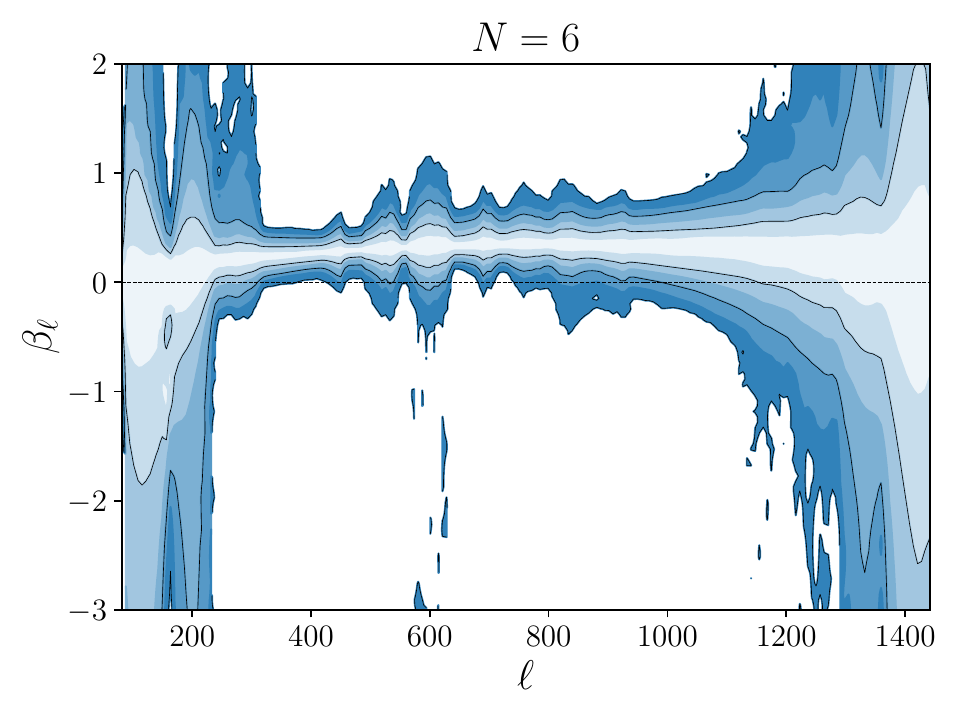}
    
    \caption{Reconstruction of the angular isotropic cosmic birefringence angle plotted using iso-probability credibility intervals in the $(\beta_\ell, \ell)$ plane, with their masses converted to $\sigma$-values via an inverse error function transformation, obtained from the $EB$-based estimator applied to (from top to bottom) the {\tt Commander}, {\tt NILC}, {\tt SEVEM} and {\tt SMICA} maps for $N=2$ (left panels) and $N=6$ (right panels).}
    \label{fig:reconstruction_EB_ne4_1}
\end{figure}

\begin{figure}[h!]
    \centering
    \includegraphics{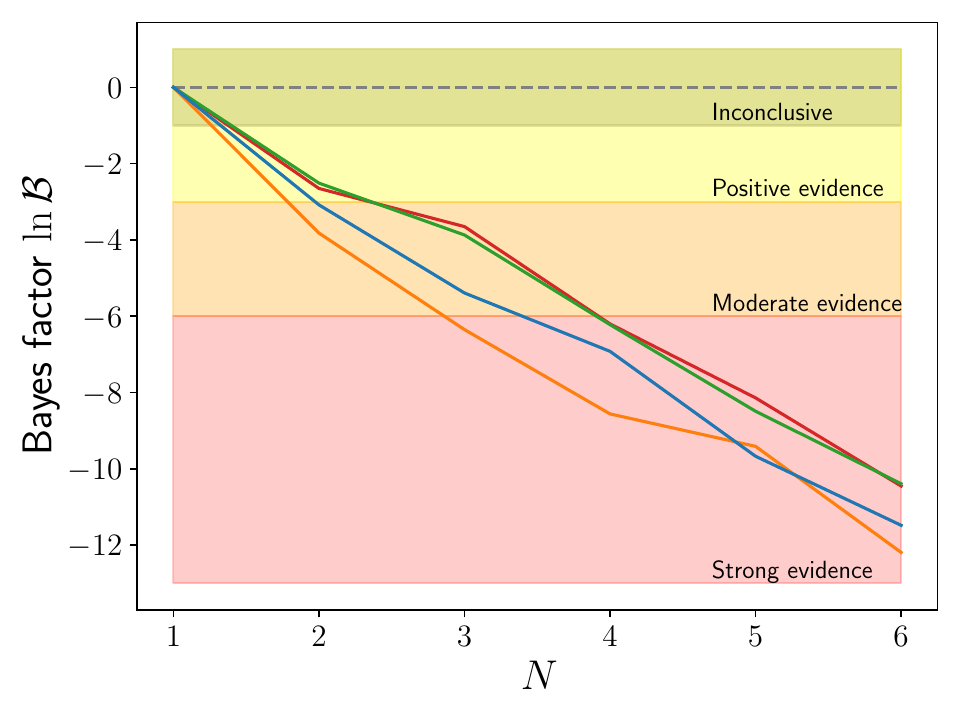}    
    \caption{Bayesian evidence as a function of number of nodes $N$ for the isotropic cosmic birefringence angle reconstruction.}
    \label{fig:reconstruction_Bayes}
\end{figure}

\Cref{fig:reconstruction_EB_ne4_1} shows the evolution of the reconstructed function as the number of nodes changes with respect to the case considered in \cref{sec:Nonparametric} for $N=4$, with the shaded regions representing the credible intervals derived from the posterior samples.
The results show how increasing the number of nodes improves the ability of the reconstruction to capture fine features of the data, while also highlighting the diminishing returns as the number of nodes becomes excessive, leading to potential overfitting. This is also described by the worsening of the Bayes factor as the number of nodes increases; see \cref{fig:reconstruction_Bayes}. 
The result agrees well with the best fit obtained with the power-law template showed in \cref{table:best_fit_couples,fig:EB_scatter_plot}.

\acknowledgments
We thank Patricia Diego-Palazuelos for her valuable feedback on the draft version of the manuscript. We acknowledge financial support from the INFN InDark initiative and from the COSMOS network through the ASI (Italian Space Agency) Grants 2016-24-H.0, 2016-24-H.1-2018, 2020-9-HH.0 (participation in LiteBIRD phase A). SSS acknowledges that this publication was produced while attending the PhD program in Space Science and Technology at the University of Trento, Cycle XXXVIII, with the support of a scholarship co-financed by the Ministerial Decree no. 351 of 9th April 2022, based on the NRRP - funded by the European Union - NextGenerationEU - Mission 4 "Education and Research", Component 2 "From Research to Business", Investment 3.3. SSS acknowledges that this publication is based upon work from COST Action CA21136 – “Addressing observational tensions in cosmology with systematics and fundamental physics (CosmoVerse)”, supported by COST (European Cooperation in Science and Technology). AG, PN and SP acknowledge support by the MUR PRIN2022 Project “BROWSEPOL: Beyond standaRd mOdel With coSmic microwavE background POLarization”-2022EJNZ53 financed by the European Union - Next Generation EU.

\paragraph{Note added.}
While we were completing our manuscript, we learned of the decision by DOE and NSF to no longer support the CMB-S4 Project. Nevertheless, we decided to keep our CMB-S4 forecast in our analysis.


 \bibliographystyle{JHEP}
 \bibliography{biblio.bib}

\providecommand{\href}[2]{#2}\begingroup\raggedright\begin{thebibliography}{10}

\bibitem{Carroll:1989vb}
S.M.~Carroll, G.B.~Field and R.~Jackiw, \emph{{Limits on a Lorentz and Parity Violating Modification of Electrodynamics}}, \href{https://doi.org/10.1103/PhysRevD.41.1231}{\emph{Phys. Rev. D} {\bfseries 41} (1990) 1231}.

\bibitem{Carroll:1991zs}
S.M.~Carroll and G.B.~Field, \emph{{The Einstein equivalence principle and the polarization of radio galaxies}}, \href{https://doi.org/10.1103/PhysRevD.43.3789}{\emph{Phys. Rev. D} {\bfseries 43} (1991) 3789}.

\bibitem{Harari:1992ea}
D.~Harari and P.~Sikivie, \emph{{Effects of a Nambu-Goldstone boson on the polarization of radio galaxies and the cosmic microwave background}}, \href{https://doi.org/10.1016/0370-2693(92)91363-E}{\emph{Phys. Lett. B} {\bfseries 289} (1992) 67}.

\bibitem{Carroll:1998zi}
S.M.~Carroll, \emph{{Quintessence and the rest of the world}}, \href{https://doi.org/10.1103/PhysRevLett.81.3067}{\emph{Phys. Rev. Lett.} {\bfseries 81} (1998) 3067} [\href{https://arxiv.org/abs/astro-ph/9806099}{{\ttfamily astro-ph/9806099}}].

\bibitem{Ni:1977zz}
W.-T.~Ni, \emph{{Equivalence Principles and Electromagnetism}}, \href{https://doi.org/10.1103/PhysRevLett.38.301}{\emph{Phys. Rev. Lett.} {\bfseries 38} (1977) 301}.

\bibitem{Ni:2016dvq}
W.-T.~Ni, \emph{{A nonmetric theory of gravity}}, \href{https://doi.org/10.1142/S0218271816400174}{\emph{Int. J. Mod. Phys. D} {\bfseries 25} (2016) 1640017}.

\bibitem{Komatsu:2022nvu}
E.~Komatsu, \emph{{New physics from the polarized light of the cosmic microwave background}}, \href{https://doi.org/10.1038/s42254-022-00452-4}{\emph{Nature Rev. Phys.} {\bfseries 4} (2022) 452} [\href{https://arxiv.org/abs/2202.13919}{{\ttfamily 2202.13919}}].

\bibitem{Lue:1998mq}
A.~Lue, L.-M.~Wang and M.~Kamionkowski, \emph{{Cosmological signature of new parity violating interactions}}, \href{https://doi.org/10.1103/PhysRevLett.83.1506}{\emph{Phys. Rev. Lett.} {\bfseries 83} (1999) 1506} [\href{https://arxiv.org/abs/astro-ph/9812088}{{\ttfamily astro-ph/9812088}}].

\bibitem{Feng:2004mq}
B.~Feng, H.~Li, M.-z.~Li and X.-m.~Zhang, \emph{{Gravitational leptogenesis and its signatures in CMB}}, \href{https://doi.org/10.1016/j.physletb.2005.06.009}{\emph{Phys. Lett. B} {\bfseries 620} (2005) 27} [\href{https://arxiv.org/abs/hep-ph/0406269}{{\ttfamily hep-ph/0406269}}].

\bibitem{Liu:2006uh}
G.-C.~Liu, S.~Lee and K.-W.~Ng, \emph{{Effect on cosmic microwave background polarization of coupling of quintessence to pseudoscalar formed from the electromagnetic field and its dual}}, \href{https://doi.org/10.1103/PhysRevLett.97.161303}{\emph{Phys. Rev. Lett.} {\bfseries 97} (2006) 161303} [\href{https://arxiv.org/abs/astro-ph/0606248}{{\ttfamily astro-ph/0606248}}].

\bibitem{Finelli:2008jv}
F.~Finelli and M.~Galaverni, \emph{{Rotation of Linear Polarization Plane and Circular Polarization from Cosmological Pseudo-Scalar Fields}}, \href{https://doi.org/10.1103/PhysRevD.79.063002}{\emph{Phys. Rev. D} {\bfseries 79} (2009) 063002} [\href{https://arxiv.org/abs/0802.4210}{{\ttfamily 0802.4210}}].

\bibitem{Fedderke:2019ajk}
M.A.~Fedderke, P.W.~Graham and S.~Rajendran, \emph{{Axion Dark Matter Detection with CMB Polarization}}, \href{https://doi.org/10.1103/PhysRevD.100.015040}{\emph{Phys. Rev. D} {\bfseries 100} (2019) 015040} [\href{https://arxiv.org/abs/1903.02666}{{\ttfamily 1903.02666}}].

\bibitem{Galaverni:2023zhv}
M.~Galaverni, F.~Finelli and D.~Paoletti, \emph{{Redshift evolution of cosmic birefringence in CMB anisotropies}}, \href{https://doi.org/10.1103/PhysRevD.107.083529}{\emph{Phys. Rev. D} {\bfseries 107} (2023) 083529} [\href{https://arxiv.org/abs/2301.07971}{{\ttfamily 2301.07971}}].

\bibitem{Planck:2016soo}
{\scshape Planck} collaboration, \emph{{Planck intermediate results. XLIX. Parity-violation constraints from polarization data}}, \href{https://doi.org/10.1051/0004-6361/201629018}{\emph{Astron. Astrophys.} {\bfseries 596} (2016) A110} [\href{https://arxiv.org/abs/1605.08633}{{\ttfamily 1605.08633}}].

\bibitem{Keating:2012ge}
B.G.~Keating, M.~Shimon and A.P.S.~Yadav, \emph{{Self-Calibration of CMB Polarization Experiments}}, \href{https://doi.org/10.1088/2041-8205/762/2/L23}{\emph{Astrophys. J. Lett.} {\bfseries 762} (2012) L23} [\href{https://arxiv.org/abs/1211.5734}{{\ttfamily 1211.5734}}].

\bibitem{Minami:2019ruj}
Y.~Minami, H.~Ochi, K.~Ichiki, N.~Katayama, E.~Komatsu and T.~Matsumura, \emph{{Simultaneous determination of the cosmic birefringence and miscalibrated polarization angles from CMB experiments}}, \href{https://doi.org/10.1093/ptep/ptz079}{\emph{PTEP} {\bfseries 2019} (2019) 083E02} [\href{https://arxiv.org/abs/1904.12440}{{\ttfamily 1904.12440}}].

\bibitem{Minami:2020fin}
Y.~Minami and E.~Komatsu, \emph{{Simultaneous determination of the cosmic birefringence and miscalibrated polarization angles II: Including cross frequency spectra}}, \href{https://doi.org/10.1093/ptep/ptaa130}{\emph{PTEP} {\bfseries 2020} (2020) 103E02} [\href{https://arxiv.org/abs/2006.15982}{{\ttfamily 2006.15982}}].

\bibitem{Minami:2020odp}
Y.~Minami and E.~Komatsu, \emph{{New Extraction of the Cosmic Birefringence from the Planck 2018 Polarization Data}}, \href{https://doi.org/10.1103/PhysRevLett.125.221301}{\emph{Phys. Rev. Lett.} {\bfseries 125} (2020) 221301} [\href{https://arxiv.org/abs/2011.11254}{{\ttfamily 2011.11254}}].

\bibitem{Diego-Palazuelos:2022dsq}
P.~Diego-Palazuelos et~al., \emph{{Cosmic Birefringence from the Planck Data Release 4}}, \href{https://doi.org/10.1103/PhysRevLett.128.091302}{\emph{Phys. Rev. Lett.} {\bfseries 128} (2022) 091302} [\href{https://arxiv.org/abs/2201.07682}{{\ttfamily 2201.07682}}].

\bibitem{Diego_Palazuelos_2023}
P.~Diego-Palazuelos, E.~Martínez-González, P.~Vielva, R.~Barreiro, M.~Tristram, E.~de~la Hoz et~al., \emph{Robustness of cosmic birefringence measurement against galactic foreground emission and instrumental systematics}, \href{https://doi.org/10.1088/1475-7516/2023/01/044}{\emph{Journal of Cosmology and Astroparticle Physics} {\bfseries 2023} (2023) 044}.

\bibitem{Eskilt:2022cff}
J.R.~Eskilt and E.~Komatsu, \emph{{Improved constraints on cosmic birefringence from the WMAP and Planck cosmic microwave background polarization data}}, \href{https://doi.org/10.1103/PhysRevD.106.063503}{\emph{Phys. Rev. D} {\bfseries 106} (2022) 063503} [\href{https://arxiv.org/abs/2205.13962}{{\ttfamily 2205.13962}}].

\bibitem{Gruppuso:2025ywx}
A.~Gruppuso and S.~di~Serego~Alighieri, \emph{{Cosmic Polarisation Rotation from CMB Data: a Review for GR110}},  \href{https://arxiv.org/abs/2502.07743}{{\ttfamily 2502.07743}}.

\bibitem{ACT:2025fju}
{\scshape ACT} collaboration, \emph{{The Atacama Cosmology Telescope: DR6 Power Spectra, Likelihoods and $\Lambda$CDM Parameters}},  \href{https://arxiv.org/abs/2503.14452}{{\ttfamily 2503.14452}}.

\bibitem{Sullivan:2025btc}
R.M.~Sullivan, A.~Abghari, P.~Diego-Palazuelos, L.T.~Hergt and D.~Scott, \emph{{Planck PR4 (NPIPE) map-space cosmic birefringence}},  \href{https://arxiv.org/abs/2502.07654}{{\ttfamily 2502.07654}}.

\bibitem{COMPACT:2024cud}
{\scshape COMPACT} collaboration, \emph{{Cosmic topology. Part IIIa. Microwave background parity violation without parity-violating microphysics}}, \href{https://doi.org/10.1088/1475-7516/2024/11/020}{\emph{JCAP} {\bfseries 11} (2024) 020} [\href{https://arxiv.org/abs/2407.09400}{{\ttfamily 2407.09400}}].

\bibitem{Nakatsuka:2022epj}
H.~Nakatsuka, T.~Namikawa and E.~Komatsu, \emph{{Is cosmic birefringence due to dark energy or dark matter? A tomographic approach}}, \href{https://doi.org/10.1103/PhysRevD.105.123509}{\emph{Phys. Rev. D} {\bfseries 105} (2022) 123509} [\href{https://arxiv.org/abs/2203.08560}{{\ttfamily 2203.08560}}].

\bibitem{Greco:2024oie}
A.~Greco, N.~Bartolo and A.~Gruppuso, \emph{{A new solution for the observed isotropic cosmic birefringence angle and its implications for the anisotropic counterpart through a Boltzmann approach}}, \href{https://doi.org/10.1088/1475-7516/2024/10/028}{\emph{JCAP} {\bfseries 10} (2024) 028} [\href{https://arxiv.org/abs/2401.07079}{{\ttfamily 2401.07079}}].

\bibitem{Fujita:2020ecn}
T.~Fujita, K.~Murai, H.~Nakatsuka and S.~Tsujikawa, \emph{{Detection of isotropic cosmic birefringence and its implications for axionlike particles including dark energy}}, \href{https://doi.org/10.1103/PhysRevD.103.043509}{\emph{Phys. Rev. D} {\bfseries 103} (2021) 043509} [\href{https://arxiv.org/abs/2011.11894}{{\ttfamily 2011.11894}}].

\bibitem{Yin:2023srb}
L.~Yin, J.~Kochappan, T.~Ghosh and B.-H.~Lee, \emph{{Is cosmic birefringence model-dependent?}}, \href{https://doi.org/10.1088/1475-7516/2023/10/007}{\emph{JCAP} {\bfseries 10} (2023) 007} [\href{https://arxiv.org/abs/2305.07937}{{\ttfamily 2305.07937}}].

\bibitem{Kochappan:2024jyf}
J.~Kochappan, L.~Yin, B.-H.~Lee and T.~Ghosh, \emph{{Observational evidence for Early Dark Energy as a unified explanation for Cosmic Birefringence and the Hubble tension}},  \href{https://arxiv.org/abs/2408.09521}{{\ttfamily 2408.09521}}.

\bibitem{Greco:2022xwj}
A.~Greco, N.~Bartolo and A.~Gruppuso, \emph{{Probing Axions through Tomography of Anisotropic Cosmic Birefringence}}, \href{https://doi.org/10.1088/1475-7516/2023/05/026}{\emph{JCAP} {\bfseries 05} (2023) 026} [\href{https://arxiv.org/abs/2211.06380}{{\ttfamily 2211.06380}}].

\bibitem{Eskilt:2023nxm}
J.R.~Eskilt, L.~Herold, E.~Komatsu, K.~Murai, T.~Namikawa and F.~Naokawa, \emph{{Constraints on Early Dark Energy from Isotropic Cosmic Birefringence}}, \href{https://doi.org/10.1103/PhysRevLett.131.121001}{\emph{Phys. Rev. Lett.} {\bfseries 131} (2023) 121001} [\href{https://arxiv.org/abs/2303.15369}{{\ttfamily 2303.15369}}].

\bibitem{Namikawa:2025sft}
T.~Namikawa, K.~Murai and F.~Naokawa, \emph{{Planck Constraints on Axion-Like Particles through Isotropic Cosmic Birefringence}},  \href{https://arxiv.org/abs/2506.20824}{{\ttfamily 2506.20824}}.

\bibitem{Planck:2018nkj}
{\scshape Planck} collaboration, \emph{{Planck 2018 results. I. Overview and the cosmological legacy of Planck}}, \href{https://doi.org/10.1051/0004-6361/201833880}{\emph{Astron. Astrophys.} {\bfseries 641} (2020) A1} [\href{https://arxiv.org/abs/1807.06205}{{\ttfamily 1807.06205}}].

\bibitem{Handley:2015fda}
W.J.~Handley, M.P.~Hobson and A.N.~Lasenby, \emph{{PolyChord: nested sampling for cosmology}}, \href{https://doi.org/10.1093/mnrasl/slv047}{\emph{Mon. Not. Roy. Astron. Soc.} {\bfseries 450} (2015) L61} [\href{https://arxiv.org/abs/1502.01856}{{\ttfamily 1502.01856}}].

\bibitem{Handley:2015vkr}
W.J.~Handley, M.P.~Hobson and A.N.~Lasenby, \emph{{polychord: next-generation nested sampling}}, \href{https://doi.org/10.1093/mnras/stv1911}{\emph{Mon. Not. Roy. Astron. Soc.} {\bfseries 453} (2015) 4385} [\href{https://arxiv.org/abs/1506.00171}{{\ttfamily 1506.00171}}].

\bibitem{Raffaelli:2025kew}
A.~{Raffaelli} and M.~{Ballardini}, \emph{{Knot reconstruction of the scalar primordial power spectrum with Planck, ACT, and SPT CMB data}}, \href{https://doi.org/10.48550/arXiv.2503.10609}{\emph{arXiv e-prints} (2025) arXiv:2503.10609} [\href{https://arxiv.org/abs/2503.10609}{{\ttfamily 2503.10609}}].

\bibitem{Eriksen:2005dr}
H.K.~Eriksen et~al., \emph{{CMB component separation by parameter estimation}}, \href{https://doi.org/10.1086/500499}{\emph{Astrophys. J.} {\bfseries 641} (2006) 665} [\href{https://arxiv.org/abs/astro-ph/0508268}{{\ttfamily astro-ph/0508268}}].

\bibitem{Eriksen:2007mx}
H.K.~Eriksen, J.B.~Jewell, C.~Dickinson, A.J.~Banday, K.M.~Gorski and C.R.~Lawrence, \emph{{Joint Bayesian component separation and CMB power spectrum estimation}}, \href{https://doi.org/10.1086/525277}{\emph{Astrophys. J.} {\bfseries 676} (2008) 10} [\href{https://arxiv.org/abs/0709.1058}{{\ttfamily 0709.1058}}].

\bibitem{Delabrouille:2008qd}
J.~Delabrouille, J.F.~Cardoso, M.L.~Jeune, M.~Betoule, G.~Fay and F.~Guilloux, \emph{{A full sky, low foreground, high resolution CMB map from WMAP}}, \href{https://doi.org/10.1051/0004-6361:200810514}{\emph{Astron. Astrophys.} {\bfseries 493} (2009) 835} [\href{https://arxiv.org/abs/0807.0773}{{\ttfamily 0807.0773}}].

\bibitem{Martinez-Gonzalez:2003abe}
E.~Martinez-Gonzalez, J.M.~Diego, P.~Vielva and J.~Silk, \emph{{CMB power spectrum estimation and map reconstruction with the expectation - Maximization algorithm}}, \href{https://doi.org/10.1046/j.1365-2966.2003.06885.x}{\emph{Mon. Not. Roy. Astron. Soc.} {\bfseries 345} (2003) 1101} [\href{https://arxiv.org/abs/astro-ph/0302094}{{\ttfamily astro-ph/0302094}}].

\bibitem{Delabrouille:2002kz}
J.~Delabrouille, J.F.~Cardoso and G.~Patanchon, \emph{{Multi-detector multi-component spectral matching and applications for CMB data analysis}}, \href{https://doi.org/10.1111/j.1365-2966.2003.07069.x}{\emph{Mon. Not. Roy. Astron. Soc.} {\bfseries 346} (2003) 1089} [\href{https://arxiv.org/abs/astro-ph/0211504}{{\ttfamily astro-ph/0211504}}].

\bibitem{Cardoso:2008qt}
J.-F.~Cardoso, M.~Martin, J.~Delabrouille, M.~Betoule and G.~Patanchon, \emph{{Component separation with flexible models. Application to the separation of astrophysical emissions}},  \href{https://arxiv.org/abs/0803.1814}{{\ttfamily 0803.1814}}.

\bibitem{Leach:2008fi}
S.M.~Leach et~al., \emph{{Component separation methods for the Planck mission}}, \href{https://doi.org/10.1051/0004-6361:200810116}{\emph{Astron. Astrophys.} {\bfseries 491} (2008) 597} [\href{https://arxiv.org/abs/0805.0269}{{\ttfamily 0805.0269}}].

\bibitem{Planck:2018yye}
{\scshape Planck} collaboration, \emph{{Planck 2018 results. IV. Diffuse component separation}}, \href{https://doi.org/10.1051/0004-6361/201833881}{\emph{Astron. Astrophys.} {\bfseries 641} (2020) A4} [\href{https://arxiv.org/abs/1807.06208}{{\ttfamily 1807.06208}}].

\bibitem{Bortolami:2022whx}
M.~Bortolami, M.~Billi, A.~Gruppuso, P.~Natoli and L.~Pagano, \emph{{Planck constraints on cross-correlations between anisotropic cosmic birefringence and CMB polarization}}, \href{https://doi.org/10.1088/1475-7516/2022/09/075}{\emph{JCAP} {\bfseries 09} (2022) 075} [\href{https://arxiv.org/abs/2206.01635}{{\ttfamily 2206.01635}}].

\bibitem{2010A&A...520A...1T}
J.A.~{Tauber}, N.~{Mandolesi}, J.L.~{Puget}, T.~{Banos}, M.~{Bersanelli}, F.R.~{Bouchet} et~al., \emph{{Planck pre-launch status: The Planck mission}}, \href{https://doi.org/10.1051/0004-6361/200912983}{\emph{AA} {\bfseries 520} (2010) A1}.

\bibitem{Planck:2015txa}
{\scshape Planck} collaboration, \emph{{Planck 2015 results. XII. Full Focal Plane simulations}}, \href{https://doi.org/10.1051/0004-6361/201527103}{\emph{Astron. Astrophys.} {\bfseries 594} (2016) A12} [\href{https://arxiv.org/abs/1509.06348}{{\ttfamily 1509.06348}}].

\bibitem{Planck:2018lkk}
{\scshape Planck} collaboration, \emph{{Planck 2018 results. III. High Frequency Instrument data processing and frequency maps}}, \href{https://doi.org/10.1051/0004-6361/201832909}{\emph{Astron. Astrophys.} {\bfseries 641} (2020) A3} [\href{https://arxiv.org/abs/1807.06207}{{\ttfamily 1807.06207}}].

\bibitem{Gorski:2004by}
K.M.~G\'orski, E.~Hivon, A.J.~Banday, B.D.~Wandelt, F.K.~Hansen, M.~Reinecke et~al., \emph{{HEALPix - A Framework for high resolution discretization, and fast analysis of data distributed on the sphere}}, \href{https://doi.org/10.1086/427976}{\emph{Astrophys. J.} {\bfseries 622} (2005) 759} [\href{https://arxiv.org/abs/astro-ph/0409513}{{\ttfamily astro-ph/0409513}}].

\bibitem{Alonso:2018jzx}
{\scshape LSST Dark Energy Science} collaboration, \emph{{A unified pseudo-$C_\ell$ framework}}, \href{https://doi.org/10.1093/mnras/stz093}{\emph{Mon. Not. Roy. Astron. Soc.} {\bfseries 484} (2019) 4127} [\href{https://arxiv.org/abs/1809.09603}{{\ttfamily 1809.09603}}].

\bibitem{Hivon:2001jp}
E.~Hivon, K.M.~Gorski, C.B.~Netterfield, B.P.~Crill, S.~Prunet and F.~Hansen, \emph{{Master of the cosmic microwave background anisotropy power spectrum: a fast method for statistical analysis of large and complex cosmic microwave background data sets}}, \href{https://doi.org/10.1086/338126}{\emph{Astrophys. J.} {\bfseries 567} (2002) 2} [\href{https://arxiv.org/abs/astro-ph/0105302}{{\ttfamily astro-ph/0105302}}].

\bibitem{Polenta:2004qs}
G.~Polenta, D.~Marinucci, A.~Balbi, P.~de~Bernardis, E.~Hivon, S.~Masi et~al., \emph{{Unbiased estimation of angular power spectrum}}, \href{https://doi.org/10.1088/1475-7516/2005/11/001}{\emph{JCAP} {\bfseries 11} (2005) 001} [\href{https://arxiv.org/abs/astro-ph/0402428}{{\ttfamily astro-ph/0402428}}].

\bibitem{Lewis:2001hp}
A.~Lewis, A.~Challinor and N.~Turok, \emph{{Analysis of CMB polarization on an incomplete sky}}, \href{https://doi.org/10.1103/PhysRevD.65.023505}{\emph{Phys. Rev. D} {\bfseries 65} (2002) 023505} [\href{https://arxiv.org/abs/astro-ph/0106536}{{\ttfamily astro-ph/0106536}}].

\bibitem{Bunn:2002df}
E.F.~Bunn, M.~Zaldarriaga, M.~Tegmark and A.~de~Oliveira-Costa, \emph{{E/B decomposition of finite pixelized CMB maps}}, \href{https://doi.org/10.1103/PhysRevD.67.023501}{\emph{Phys. Rev. D} {\bfseries 67} (2003) 023501} [\href{https://arxiv.org/abs/astro-ph/0207338}{{\ttfamily astro-ph/0207338}}].

\bibitem{Grain:2009wq}
J.~Grain, M.~Tristram and R.~Stompor, \emph{{Polarized CMB spectrum estimation using the pure pseudo cross-spectrum approach}}, \href{https://doi.org/10.1103/PhysRevD.79.123515}{\emph{Phys. Rev. D} {\bfseries 79} (2009) 123515} [\href{https://arxiv.org/abs/0903.2350}{{\ttfamily 0903.2350}}].

\bibitem{QUaD:2008ado}
{\scshape QUaD} collaboration, \emph{{Parity Violation Constraints Using Cosmic Microwave Background Polarization Spectra from 2006 and 2007 Observations by the QUaD Polarimeter}}, \href{https://doi.org/10.1103/PhysRevLett.102.161302}{\emph{Phys. Rev. Lett.} {\bfseries 102} (2009) 161302} [\href{https://arxiv.org/abs/0811.0618}{{\ttfamily 0811.0618}}].

\bibitem{Gruppuso:2016nhj}
A.~Gruppuso, G.~Maggio, D.~Molinari and P.~Natoli, \emph{{A note on the birefringence angle estimation in CMB data analysis}}, \href{https://doi.org/10.1088/1475-7516/2016/05/020}{\emph{JCAP} {\bfseries 05} (2016) 020} [\href{https://arxiv.org/abs/1604.05202}{{\ttfamily 1604.05202}}].

\bibitem{Hartlap:2006kj}
J.~Hartlap, P.~Simon and P.~Schneider, \emph{{Why your model parameter confidences might be too optimistic: Unbiased estimation of the inverse covariance matrix}}, \href{https://doi.org/10.1051/0004-6361:20066170}{\emph{Astron. Astrophys.} {\bfseries 464} (2007) 399} [\href{https://arxiv.org/abs/astro-ph/0608064}{{\ttfamily astro-ph/0608064}}].

\bibitem{Millea:2018bko}
M.~Millea and F.~Bouchet, \emph{{Cosmic Microwave Background Constraints in Light of Priors Over Reionization Histories}}, \href{https://doi.org/10.1051/0004-6361/201833288}{\emph{Astron. Astrophys.} {\bfseries 617} (2018) A96} [\href{https://arxiv.org/abs/1804.08476}{{\ttfamily 1804.08476}}].

\bibitem{Handley:2019fll}
W.J.~Handley, A.N.~Lasenby, H.V.~Peiris and M.P.~Hobson, \emph{{Bayesian inflationary reconstructions from Planck 2018 data}}, \href{https://doi.org/10.1103/PhysRevD.100.103511}{\emph{Phys. Rev. D} {\bfseries 100} (2019) 103511} [\href{https://arxiv.org/abs/1908.00906}{{\ttfamily 1908.00906}}].

\bibitem{fgivenx}
W.~Handley, \emph{fgivenx: Functional posterior plotter}, \href{https://doi.org/10.21105/joss.00849}{\emph{The Journal of Open Source Software} {\bfseries 3} (2018) }.

\bibitem{LiteBird:2022}
{LiteBIRD Collaboration}, E.~Allys, K.~Arnold, J.~Aumont, R.~Aurlien, S.~Azzoni et~al., \emph{Probing cosmic inflation with the litebird cosmic microwave background polarization survey}, \href{https://doi.org/10.1093/ptep/ptac150}{\emph{Progress of Theoretical and Experimental Physics} {\bfseries 2023} (2022) 042F01} [\href{https://arxiv.org/abs/https://academic.oup.com/ptep/article-pdf/2023/4/042F01/50954267/ptac150.pdf}{{\ttfamily https://academic.oup.com/ptep/article-pdf/2023/4/042F01/50954267/ptac150.pdf}}].

\bibitem{delaHoz:2025uae}
E.~de~la Hoz et~al., \emph{{LiteBIRD Science Goals and Forecasts: constraining isotropic cosmic birefringence}},  \href{https://arxiv.org/abs/2503.22322}{{\ttfamily 2503.22322}}.

\bibitem{Ade_2019}
P.~Ade, J.~Aguirre, Z.~Ahmed, S.~Aiola, A.~Ali, D.~Alonso et~al., \emph{The simons observatory: science goals and forecasts}, \href{https://doi.org/10.1088/1475-7516/2019/02/056}{\emph{Journal of Cosmology and Astroparticle Physics} {\bfseries 2019} (2019) 056}.

\bibitem{SimonsObservatory:2025wwn}
{\scshape Simons Observatory} collaboration, \emph{{The Simons Observatory: Science Goals and Forecasts for the Enhanced Large Aperture Telescope}},  \href{https://arxiv.org/abs/2503.00636}{{\ttfamily 2503.00636}}.

\bibitem{Abazajian_2022}
K.~Abazajian, G.E.~Addison, P.~Adshead, Z.~Ahmed, D.~Akerib, A.~Ali et~al., \emph{Cmb-s4: Forecasting constraints on primordial gravitational waves}, \href{https://doi.org/10.3847/1538-4357/ac1596}{\emph{The Astrophysical Journal} {\bfseries 926} (2022) 54}.

\bibitem{CMBs4:2024}
E.~Schiappucci, S.~Raghunathan, C.~To, F.~Bianchini, C.L.~Reichardt, N.~Battaglia et~al., \emph{Constraining cosmological parameters using the pairwise kinematic sunyaev-zel'dovich effect with cmb-s4 and future galaxy cluster surveys},  2024.

\end{thebibliography}\endgroup

\end{document}